\documentclass{imsart}

\usepackage{orcidlink,thumbpdf,lmodern}
\usepackage{listings}%for overleaf verbatim of R code
\usepackage[utf8]{inputenc}
\usepackage{natbib}
\usepackage{bm}
\usepackage{amsmath}
\usepackage{amssymb}
\usepackage{mathptmx}
\usepackage{tikz} % Required for flow chart
\usetikzlibrary{shapes,arrows}

\newcommand{\mbf}[1]{\mathbf{#1}}
\newcommand{\appsim}{\stackrel{\mathrm{a}}{\sim}}
\usepackage{bm}
\usepackage[linesnumbered,ruled]{algorithm2e}

\begin{document}

\begin{frontmatter}

\title{Uncertainty Quantification for Multi-level Models Using the Survey-Weighted Pseudo-Posterior}
\runtitle{Bayesian Models for Survey Data}

\begin{aug}
\author{\fnms{Matthew R.} \snm{Williams}\thanksref{addr1}}
\and
\author{\fnms{F. Hunter} \snm{McGuire}\thanksref{addr2}}
\and
\author{\fnms{Terrance D.} \snm{Savitsky}\thanksref{addr3}}

\runauthor{Williams, McGuire \and Savitsky}

\address[addr1]{RTI International
    \href{mailto:mrwilliams@rti.org}{mrwilliams@rti.org}
}
\address[addr2]{PicnicHealth}
\address[addr3]{U.S. Bureau of Labor Statistics}

\end{aug}

%\runtitle{}  %use if necessary to change auto-generated
%\runauthor{} %use if necessary to change auto-generated

%\dedicated{}

\begin{abstract} % avoid notations; avoid citations; no exceeding 250 words
Parameter estimation and inference from complex survey samples typically focuses on global model parameters whose estimators have asymptotic properties, such as from fixed effects regression models.  The central challenge is to both mitigate bias induced from potentially unbalanced samples and to incorporate adjustments for differences in effective sample size  to get correct variance and interval estimates. We present a motivating example of Bayesian inference for a multi-level or mixed effects model in which estimates of both the local parameters (e.g. group level random effects) and the global parameters need to be adjusted for the complex sampling design. We evaluate the limitations of the survey-weighted pseudo-posterior and an existing automated post-processing method to improve the uncertainty quantification. We propose modifications to the automated process and demonstrate their improvements for multi-level models via a simulation study and a motivating example from the National Survey on Drug Use and Health. Reproduction examples are available from the authors and the updated R package is available via github: \href{https://github.com/RyanHornby/csSampling}{https://github.com/RyanHornby/csSampling}
\end{abstract}


\end{frontmatter}
\section{Introduction}\label{introduction} 
Linear regression models are the backbone of statistical analyses for a variety of fields including the social, economic, behavioral, and health sciences. Furthermore, multi-level regression models which allow for hierarchical specification of sources of randomness (e.g. group vs. individual level variations) have become ubiquitous. Multipurpose statistical modeling packages such as {\bf brms} \citep{brms} provide a powerful suite of models and features for performing Bayesian analysis for multilevel models, while using syntax closely related to base linear regression. However, key social and economic data sources are collected not as simple random samples, but through complex sampling designs that include sampling of clusters and unequal probabilities of selection. Current software packages can be coerced to produce estimates for these models that are asymptotically consistent by including survey weights. However, the resulting standard errors and credible intervals will not produce correct coverage (e.g. a 95\% interval will not contain the true value 95\% of the time).

Our motivation is to extend the use of variance component and other hierarchical models to the case where data are collected under a complex survey sampling design associated with a finite population. For simple regression models, well established theory and tools exist (e.g. the {\bf survey} package \citep{Rsurvey}), allowing for the incorporation of features of the survey sampling design (unequal probabilities of selection, stratification, and clustering) to provide asymptotically correct point and interval estimation. 
%Our goal is to address more complicated, hierarchical statistical models for complex survey data that are not readily handled by current software.
%
\cite{WilliamsISR21} develop the basic framework Bayesian analysis via a post-estimation adjustment of the survey-weighted posterior via a sandwich adjustment to the covariance matrix. However, their framework only directly applies to so-called global parameters which asymptotically have increasing information about them from the data. For example, we may expect global parameter variances to decrease proportional to the sample size $\propto n^{-1}$. 
While Bayesian models may capture an arbitrarily complex dependence or covariance structure, this complexity is typically achieved with local parameters dependent on a small portion of the sample observations. These local features might have variances proportional to a slower rate such as $\propto n^{-1/2}$ if we believe that these small groups would grow at a slower rate than than the total sample.  Some hyper-parameters for deeper models might have every slower rates, close to constant. Achieving asymptotic normality of the joint distribution of both global and local parameters may require much larger sample sizes than is practical. Under this setting, the normality assumption may be violated resulting in under-coverage of intervals when using the sandwich adjustment of \cite{WilliamsISR21}.   Yet, Bayesian analysts seek correct uncertainty quantification under small samples.

%For example the regression coefficients for a fixed effect linear model). In practice, Bayesian hierarchical models have many local parameters which are strongly influenced by both the data and the structure of the prior distribution. For moderate and even large sample sizes, the information from the likelihood does not grow at the same rate as that of the global parameters. Thus the practitioner is left with uncertainty about whether and how to adjust all parameters, with some on a continuum between fully global and fully local.

We demonstrate this challenge of identifying which parameters can be reasonably approximated by a normal distribution and which cannot through a simple group-level random effects model. In particular, we are motivated by variance decomposition approaches to model intersectionality effects on health outcomes \citep{10.1093/aje/kwae121}.
We consider a two-level generalized linear mixed model. In particular, we consider a logistic regression with fixed effects and random intercepts indexed by groups within the population. 
\begin{equation}\label{eq:logisticMM}
\begin{array}{rl}
    y_{ij} &\sim \ Bern(p_{ij})\\
    logit(p_{ij}) & = \ \mu_{ij} = X_{ij} \beta + \alpha_j\\
    \beta & \sim \ N(0, \Sigma)\\
    %\Sigma & \sim \ G\\
    \alpha_j & \sim \ N(0, \sigma_\alpha)\\
\end{array}
\end{equation}
This will be our main analysis model, which we will evaluate through simulation and on a large scale complex survey.

In the intersectionality setting, we have groups $j$ defined by stratifying the population by crossing several demographic variables. We need reliable estimates of both the global parameters $\beta$ and $\sigma_\alpha$ as well as the group level estimates $\alpha_j$ to produce marginal estimates of prevalence for each group $j$ averaged over the covariate values $X_{ij}$. Reliable estimates of $\beta$ will provide estimates for the main effects or `single' group effects. Reliable estimates for the group level estimates $\alpha_j$ will provide insights into how different the groups is from the sum of its main effects. For example a person with a combination of race, gender, and sexual orientation have a different expected outcome than is predicted by simply combining separate (fixed) effects from race, gender, and sex.
We also wish to approximate the variance contribution, comparing total variation in the outcomes to the contribution coming from between group variation \citep{goldstein2002}.

In Section \ref{sec:sandwich}, we review the motivation behind the survey-weighted pseudo-posterior and a post-processing sandwich adjustment to the posterior variance. In Section \ref{sec:implement}, we compare different implementation options for the automating such an adjustment. We then compare these approaches in a series of simulation studies (Section \ref{sec:sims}). We then apply our methods to the National Survey on Drug Use and Health (Section \ref{sec:NSDUH}). We conclude in Section \ref{sec:discuss} with a discussion. Additional results are included in the Appendices.

\section{Approaches to Adjusting for Survey Design}\label{sec:sandwich}
Suppose that we wish to estimate a statistical model with parameters $\theta$, where $\theta_0$ are the true unknown parameter values and $P_{\theta_{0}}$ describes the data generating mechanism given those parameters. In our motivating example, the parameters $\theta = \{\beta, \Sigma,\sigma_\alpha\}$. We wish to make inference on data $Y$ for a population $U = (1,\ldots N)$ of size $N$. However, we only have a selected sample $S = (1,\ldots, n \le N)$ drawn from the underlying population. The sampling process is governed by a distribution $P_{\nu}$, which is the joint distribution over the inclusion indicator variables $(\delta_i, \ldots \delta_N)$. The resulting joint distribution $P_{\theta_{0}}, P_{\nu}$ governs the observed sample we have.

The design $P_{\nu}$ defines the marginal probabilities of selection (or inclusion) $\pi_i = Pr(\delta_i = 1)$ as well as all higher order joint inclusions, e.g. $\pi_{ii'} = Pr(\delta_i = 1, \delta_{i'} = 1)$. A design $P_{\nu}$ is called `informative', if the distribution of $Y$ is different between the sample and the population. Informative sample designs pose two challenges for inference: (i) Estimation on the sample without accounting for unequal selection  $\pi_i$ will lead to biased estimates. (ii) Failure to account for dependence induced by sample design will lead to uncertainty estimates (standard errors and confidence intervals) that are incorrect. It may be the case that conditioning on covariates $X$, the conditional distribution of $Y$ (e.g. the relationship between $Y$ and $X$) \emph{is} the same in the sample and the population. Thus the `informativeness' of the sample depends on the outcome, and the covariates available in the sample, and the analysis model.

The first challenge can be addressed by using the survey weights $w_i \propto 1/\pi_i$. In order to approximate the population posterior
\begin{equation*}
    \pi\left(\bm{\theta}\vert \mathbf{y}\right) \propto \mathop{\prod}_{i = 1}^{N}\pi\left(y_{i}\vert \bm{\theta}\right)\pi\left(\bm{\theta}\right),
    \end{equation*}
we use the survey-weighted pseudo-posterior for the observed sample resulting in a pseudo-posterior:
\begin{equation}\label{eq:pseudopost}
    \pi^{\pi}\left(\bm{\theta}\vert \mathbf{y},\mathbf{w}\right) \propto \left[\mathop{\prod}_{i = 1}^{n}\pi\left(y_{i}\vert \bm{\theta}\right)^{w_{i}}\right]\pi\left(\bm{\theta}\right).
    \end{equation}
In practice we follow \cite{2015arXiv150707050S} by using weights scaled to sum to the sample size $\sum_i w_i = n$. This provides for a coarse adjustment to the posterior for uncertainty quantification. However, further post-hoc adjustments are often needed.

The incorporation of survey weights is a relatively minor change to the underlying model and easy to implement, especially in general modeling software which already allows for weights.
This modification leads to consistent estimation (i.e. the posterior mean converges to the true value $\theta_0$ and posterior distribution (or intervals) shrink, collapsing around the true value $\theta_0$). This consistency is with respect to the joint process of the population generation $P_{\theta_{0}}$ and the taking of samples $P_{\nu}$. The assumptions needed for consistency of the posterior fall into two classes: (i) conditions on the model and prior and (ii) conditions on the sample design. For the model, there are restrictions on the complexity (e.g. number of parameters) compared to the sample size. For the prior, there are restrictions on having enough prior mass on the true parameter (to exclude cases of very strong and very wrong priors). For the sample design, the assumptions require all individuals in the population to have a non-zero chance of being selected. Selection of individuals needs to be close to independent, with deviations from this being a small fraction, and the overall sample size needs to grow as a rate close to the population size. For more technical details, see  \cite{2015arXiv150707050S} and \cite{2018dep}. It is worth noting that the consistency for the survey-weighted pseudo-maximum likelihood estimator (MLE) has been known for many years and has been demonstrated for multiple applications with similar restrictions on the sample design but different conditions on the models. See \cite{Isaki82} for an example of an earlier reference and \cite{Han2021} for a recent comprehensive review.

The second challenge is asymptotically correct uncertainty quantification. 
%The survey-weighted pseudo-posterior is not `fully' Bayesian, in that it is not fully generative. The resulting estimated model and parameters \emph{can} be used to generate population level outcomes, but not the original sample. Thus these models are `partially' generative, because they condition on the observed sample selection $P_{\nu}$ but do not model the joint process $P_{\theta_{0}},P_{\nu}$. 
While consistent estimation for $P_{\theta_{0}}$ is possible, the models are misspecified with respect to $P_{\nu}$. These kinds of `working' models lead to misleading posterior (credible) intervals, in the sense that their frequentist coverages are incorrect (e.g. a 95\% posterior interval does not have 95\% coverage) even for increasingly larger sample sizes. See \cite{2009arXiv0911.5357R} for a closely related example for posterior inference based on composite likelihoods. For correctly specified models, we expect asymptotically that the sampling distribution of the maximum likelihood estimator and the posterior distribution will both be normal with the same mean and variance $N(\theta_0, n^{-1}H^{-1}_{\theta_0})$ (called a Bernstein-von Mises result). For a textbook level presentation of Bernstein-von Mises results for many classes of Bayesian models, see \cite{ghosal2017fundamentals}. For misspecified or `working' models such as the survey-weighted pseudo-posterior, the two will have different covariance matrices.  See \cite{kleijn2012} for more technical details. In the next section \ref{sec:covar} we define these covariance matrices for our pseudo-posterior example. %In the supplementary materials, we provide a high-level summary of the Bernstein-von Mises result from \cite{WilliamsISR21} for the setting of the pseudo-posterior.

\subsection{Asymptotic Covariances}\label{sec:covar}
We briefly review the different covariance matrices that arise from complex survey sampling.
To facilitate comparisons between these covariance matrices, we first use the following empirical distribution approximation for the joint distribution over population generation and the drawing of an informative sample that produces the observed sample.  This empirical distribution construction follows \citet{breslow:2007} and incorporates inverse inclusion probability weights, $\{1/\pi_{i}\}_{i=1,\ldots,N}$, to account for the informative sampling design \citep{2015arXiv150707050S}.
\begin{equation}
\mathbb{P}^{\pi}_{N_{\nu}} = \frac{1}{N}\mathop{\sum}_{i\in U}\frac{\delta_{i}}{\pi_{ i}}\delta\left(y_{i}\right),
\end{equation}
where $\delta\left(y_{i}\right)$ denotes the Dirac delta function, with probability mass $1$ on $y_{i}$. This construction contrasts with the usual empirical distribution, $\mathbb{P}_{N} = \frac{1}{N}\mathop{\sum}_{i\in U}\delta\left(y_{i}\right)$. Using the notation of \citet{Ghosal00convergencerates} we define expectation functionals with respect to these empirical distributions by $\mathbb{P}^{\pi}_{N}f = \frac{1}{N}\mathop{\sum}_{i=1}^{N}\frac{\delta_{i}}{\pi_{i}}f\left(y_{ i}\right)$.  Similarly, $\mathbb{P}_{N}f = \frac{1}{N}\mathop{\sum}_{i=1}^{N}f\left(y_{i}\right)$.
Starting with the Fisher Information matrix for a simple random sample:
\begin{equation*}
H_{\theta_{0}} = - \mathbb{E}_{P_{\theta_{0}}} \left[\mathbb{P}_{N} \ddot{\ell}_{\theta_{0}}\right]
= -\frac{1}{N}\mathop{\sum}_{i\in U}\mathbb{E}_{P_{\theta_{0}}}\ddot{\ell}_{\theta_{0}}(y_{i}),
\end{equation*} 
which is the negative of the expected value of the second derivative of the log-likelihood $\ddot{\ell}_{\theta_{0}}$. Under an SRS, $H_{\theta_{0}}^{-1}$ is the asymptotic variance for the MLE estimate $\hat{\theta}$.
It turns out that the survey-weighted Fisher Information matrix is the same as the SRS version, $H^{\pi}_{\theta_{0}} =  H_{\theta_{0}}$:
\begin{equation*}
 H^{\pi}_{\theta_{0}} = - \mathbb{E}_{P_{\theta_{0}}, P_{\nu}} \left[\mathbb{P}^{\pi}_{N_{\nu}} \ddot{\ell}_{\theta_{0}}\right]
 = -\frac{1}{N}\mathop{\sum}_{i\in U}\mathbb{E}_{P_{\theta_{0}}}\left[\mathbb{E}_{P_{\nu}}\frac{\delta_{ i}}{\pi_{ i}}\ddot{\ell}_{\theta_{0}}(y_i)\right] = -\frac{1}{N}\mathop{\sum}_{i\in U}\mathbb{E}_{P_{\theta_{0}}}\ddot{\ell}_{\theta_{0}}(y_{i}).
\end{equation*}

 The survey-weighted MLE has a sandwich form to its variance $H_{\theta_{0}}^{-1} J^{\pi}_{\theta_{0}} H_{\theta_{0}}^{-1}$ with
\begin{equation*}
\begin{array}{rl}
J^{\pi}_{\theta_{0}} \ = & 
\mathbb{V}_{P_{\theta_{0}},P_{\nu}}\left[\mathbb{P}^{\pi}_{N_{\nu}}\dot{\ell}_{\theta_{0}} \right]
=\mathbb{E}_{P_{\theta_{0}},P_{\nu}}\left[\left(\mathbb{P}^{\pi}_{N_{\nu}}\dot{\ell}_{\theta_{0}}\right)\left(\mathbb{P}^{\pi}_{N_{\nu}}\dot{\ell}_{\theta_{0}}\right)^{T} \right] 
\\
J_{\theta_{0}} \ = &
\mathbb{V}_{P_{\theta_{0}}}\left[\mathbb{P}_{N}\dot{\ell}_{\theta_{0}} \right] = 
\mathbb{E}_{P_{\theta_{0}}}\left[\left(\mathbb{P}_{N}\dot{\ell}_{\theta_{0}}\right)\left(\mathbb{P}_{N}\dot{\ell}_{\theta_{0}}\right)^{T} \right] 
\\
J^{\pi}_{\theta_{0}} \ \ne & J_{\theta_{0}},
\end{array}
\end{equation*}
where $J^{\pi}_{\theta_0}$ is the variance of the gradient of the log-likelihood with respect to the joint distribution of $P_{\theta_{0}},P_{\nu}$.
Under correctly specified models, $J_{\theta_0} = H_{\theta_0}$ (Bartlett's $2^{nd}$ identity) and the sandwich collapses. However for misspecified models, such as those using a pseudo-likelihood, these two matrices are distinct $J^{\pi}_{\theta_{0}} \ne H_{\theta_{0}}$, leading to the sandwich form (Godambe Information matrix).
In contrast, the survey-weighted posterior behaves as if it were taken from a simple random sample, with variance $H_{\theta_{0}}^{-1}$. The key challenge then is to adjust the pseudo-posterior such that its variance will match that of the MLE ($H_{\theta_{0}}^{-1} J^{\pi}_{\theta_{0}} H_{\theta_{0}}^{-1}$).

\subsection{Curvature Adjustments}
\cite{2009arXiv0911.5357R} propose a curvature adjustment to the pseudo-posterior sample and embed it into their MCMC procedure. However,  \cite{WilliamsISR21} perform this adjustment post-hoc, instead of inserting significant complexities into the underlying MCMC sampling process. Let $\hat{\theta}_m$ represent the sample from the pseudo-posterior for $m = 1, \ldots, M$ draws with sample mean $\bar{\theta}$.
Define the adjusted sample:
\begin{equation}\label{eq:adjustment}
\hat{\theta}^{a}_m = \left(\hat{\theta}_m - \bar{\theta}\right) R^{-1}_2 R_1  + \bar{\theta},
\end{equation}
where $R_1$ and $R_2$ are `square root' matrices such that $R'_1 R_1 = H_{\theta_{0}}^{-1} J^{\pi}_{\theta_{0}} H_{\theta_{0}}^{-1}$ and $R'_2 R_2 = H_{\theta_{0}}^{-1}$.
Asymptotically, $\hat{\theta}_m \appsim N(\theta_{0}, n^{-1} H_{\theta_{0}}^{-1})$. Then we now have $\hat{\theta}^{a}_m \appsim N(\theta_{0}, n^{-1} H_{\theta_{0}}^{-1} J^{\pi}_{\theta_{0}} H_{\theta_{0}}^{-1})$, which is the asymptotic distribution of the MLE under the pseudo-likelihood. 

In survey statistics, we often estimate a `design effect' \citep[see, for example][]{kish1995methods} which is a ratio of the (asymptotic) variance of an estimate under the complex survey design vs. its (asymptotic) variance under a simple random sample. This helps us understand the penalty or efficiency of the design (in terms of variance per fixed sample size) compared to a simple random sample. In this way, the adjustment $R^{-1}_2 R_1$ is a multivariate estimate of a design effect, providing a parameter-specific adjustment for effective sample size for variances and as well as for correlations between parameters. Adjusting the parameters simultaneously allows for us to adjust all quantities derived from adjusted parameters in downstream inference, such as predicted values. The key focus of this work is to (1) investigate whether or not both global and local parameters need an adjustment for design effect and (2) whether an adjustment can be effectively applied to both global and local parameters in an automated implementation.

In order to apply these adjustments, we must estimate the resulting variance matrices $H_\theta$ and $J^{\pi}_\theta$. 
%We will use the hybrid approach from \cite{WilliamsISR21} as implemented in the {\bf csSampling} package. We estimate the unadjusted survey-weighted posterior via the {\bf Rstan} package. We then estimate the Fisher Information $H_{\theta_{0}}$. Consistent estimates of $H_{\theta_{0}}$ are available using either the plug-in estimate $-\mathop{\sum}_{i \in S} w_i \ddot{\ell}_{\bar{\theta}}(y_{i})$ using the posterior mean $\bar{\theta}$ or the average over posterior samples $-\frac{1}{M}\mathop{\sum}_{m=1}^{M} \mathop{\sum}_{i \in S} w_i \ddot{\ell}_{\hat{\theta}_{m}}(y_{i})$. We can obtain $-\mathop{\sum}_{i \in S} w_i \ddot{\ell}_{\theta}(y_{i})$ as a function of $\theta$ by using the gradient from \texttt{grad\_log\_prob} and finite difference approaches via \texttt{optimHess} in the {\bf stats} package in R.
%
The weighted pseudo-posterior samples alone are not sufficient to estimate $J^{\pi}_{\theta_{0}}$. Instead we can only obtain estimates of $J_{\theta} = H_\theta$. To remedy this, we must find a way to use survey design-based variance estimation strategies.
To achieve this, we use a hybrid approach that is a variation on Taylor linearization (For a review of Taylor linearization methods, see \cite{binder1996}).
Instead of directly estimating the variance of $\hat{\theta}$, \cite{WilliamsISR21} target the variance of the transformed parameters:
\begin{equation}\label{eq:variance}
(\hat{\phi} - \phi_0) = H_{\theta_0} (\hat{\theta} - \theta_0) \approx \mathop{\sum}_{i \in S} w_i \dot{\ell}_{\hat{\theta}}(y_{i}) = \mathop{\sum}_{i \in S} w_i z_i (\hat{\theta}),
\end{equation}
where $Var_{P_{\theta_0}, P_{\nu}} (\hat{\phi} - \phi_0) \approx J^{\pi}_{\theta_{0}}$.
Once we estimate $J^{\pi}_{\theta_{0}}$, we back transform to solve for the variance of $\hat{\theta}$ as the sandwich $H_{\theta_{0}}^{-1} J^{\pi}_{\theta_{0}} H_{\theta_{0}}^{-1}$
In order to estimate $J^{\pi}_{\theta_{0}}$, we use a variance estimation method for the linearized term $\mathop{\sum}_{i \in S} w_i \dot{\ell}_{\hat{\theta}}(y_{i})$ using a plug-in estimate $\hat{\theta}$. We recall that a replication based approach, such as a jackknife, creates structured subsamples of the data, estimates the target of interest, and calculates the between replicate variance as an estimate of the overall variance. There are many variations of this approach \citep{Rao92}, with the key being the preservation of the sample design features (e.g. clustering, stratification, and unequal weights) across all replicates.

This hybrid variance estimation approach is implemented in {\bf csSampling}. A replication design in used to estimate $J^{\pi}_{\theta_{0}}$ but the use of the resulting sandwich adjustment from $H_{\theta_{0}}^{-1} J^{\pi}_{\theta_{0}} H_{\theta_{0}}^{-1}$ is a linearization approach. We seek to perform the intense MCMC computation to make posterior draws once, followed by a post-hoc correction to get approximately correct uncertainty quantification.

\section{Adjustment for Local Parameters}\label{sec:implement}
There are many different options for replicate designs \citep{Rao92} ranging from pure bootstrap resampling to structured balanced designs (eg. jackknife and balanced repeated replication). In practice, the results are often qualitatively similar, particularly for means and single regression coefficients.
\cite{WilliamsISR21} chose the half-sample bootstrap approach which selects half the primary sampling units with equal probability and doubles their weights, which is implemented as the default in {\bf csSampling}. For global parameters, the use of the half-sample bootstrap appears to provide adequate uncertainty estimates. However, for random effects model with a skewed population, some groups $j$ with small sample size $n_j$ may have small sample representation in several of the bootstrap replicates. That could lead to biased estimation of parameters. Instead, we investigate using a delete-a-group jack-knife approach in which one primary sampling unit (or some aggregated collection of these) is removed, while the remaining units are included. We find that this leads to larger and more stable sample sizes for small groups across all the replicates.

Using the same common delete-a-group jackknife replicates, we compare three alternatives of this basic sandwich approach which could impact the effectiveness of automatic adjustment across both global and local parameters.

\subsection{Na\"\i ve Adjustment (Baseline)}
We first describe the default approach implemented from \cite{WilliamsISR21} in the initial versions of {\bf csSampling}.
Stan allows for the extraction of the log of the posterior density and the gradient of the log posterior density. These must be available in order to implement Hamiltonian Monte Carlo (HMC) sampling. However, the gradient of the log-likelihood is not readily available. Let $\xi_{\theta} = \ell_{\theta}(y) + \log \pi_{\theta}$ be the log posterior density with gradient $\dot{\xi}_{\theta} = \dot{\ell}_{\theta}(y) + \dot{\log \pi_{\theta}}$. The simplest approach is to use these in place of the log-likelihood $\ell_{\theta}$ and $\dot{\ell}_{\theta}$. The Hessian of log posterior $H_{\xi} = - \mathbb{E}_{P_{\theta_{0}}, P_{\nu}} \left[\mathbb{P}^{\pi}_{N_{\nu}} \ddot{\xi}_{\theta_{0}}\right] = H_{\theta_{0}} + H^{0}_{\theta_{0}}$.  We assume that the curvature of the prior $H^{0}_{\theta_{0}}$ is a negligible contribution to the precision matrix $H_{\theta}$. Then for convenience we use $\dot{\xi}_{\theta}$ instead of $\dot{\ell}_{\theta}(y)$ for estimating $J^{\pi}_{\theta_{0}}$ and  $H_{\theta_{0}}$.  This is often a reasonable assumption for global parameters such as regression coefficients. However, it is a dubious assumption for local parameters and hyper-parameters.

\subsection{Using the Prior Curvature in the Adjustment}
While the Hessian of the log posterior $H_{\xi} = H_{\theta_{0}} + H^{0}_{\theta_{0}}$, the between replicate variance does not contain an extra term from the prior:
$J^{\pi}_{\xi} = \mathbb{V}_{P_{\theta_{0}},P_{\nu}}\left[\mathbb{P}^{\pi}_{N_{\nu}}\dot{\xi}_{\theta_{0}} \right] = J^{\pi}_{\theta_{0}}$. The prior density does not contain any data $y$ or sample indicators or weights. 
Then for a subset of local or hyper-parameters (e.g. $\alpha_j$), entries in $J^{\pi}_{\xi}$ will be too small compared to $H_{\xi}$, leading to an artificial reduction in scale. To compensate, we can first estimate $H^{0}$ directly and use 
$H_{\xi} = H_{\theta} + H^{0}$ which is computed directly via gradients in Stan, and we can use
$J^{\pi}_{*} = J^{\pi}_{\xi} + H^{0}$. Thus we use estimates of $H_{\xi}$ and $J^{\pi}_{*}$ in any curvature adjustments. 
In practice many Stan models, such as those created by the {\bf brms} package, have options to sample from the priors only, which allows us to construct an estimate for $H^{0}$ using the same approach as for $H_\theta$ with little additional effort.

Alternatively, we could use $H_{\xi} - H^{0}$ and $J^{\pi}_{\xi}$ instead of $H_{\xi}$ and $J^{\pi}_{*}$ for the sandwich adjustment. As noted above for global parameters, the two are asymptotically equivalent. However, for local parameters and smaller sample sizes, keeping the prior curvature $H^{0}$ provides a stabilizing effect, as the variation in some local parameters will have a large contribution from the prior.

\subsection{Using Transformations toward Normality}
For MCMC sampling, Stan transforms all model parameters to $(-\infty, \infty)$. 
For example, a variance parameter specified as $\sigma_\alpha \in [0,\infty)$ is given a log transformation to map to $(-\infty, \infty)$.
By default, {\bf csSampling} estimates the $H_\theta$ and $J^{\pi}_\theta$ and produces an adjustment on this `unconstrained' parameter space.
For variance components and other hierarchical model parameters, it is plausible that even unconstrained parameterizations in Stan (e.g. $\log(\sigma_\alpha))$ maybe still be far from normal. This has  several implications:

\begin{itemize}
    \item We can use univariate transformations to get each parameter marginally closer to a Gaussian distribution (ignoring multivariate normality).
    \item Estimates for the Hessian $H_\theta$ may be a poor estimate of the (inverse) of the posterior variance. We can consider using the posterior sample itself to get this variance since the two are equivalent asymptotically. 
    \item We expect that estimates for $J^{\pi}_\theta$ should be more robust due to the resampling approach.
\end{itemize}
We propose the following approach which automates the first two points.
\begin{itemize}
    \item Use the \cite{yeojohnson} testing and transformation approach (available in in the {\bf car} package \cite{car}).  This is a modification of the Box-Cox approach \citep{box1964analysis} that is extended to negative valued variables. We call this transformation $\psi(\theta) = \eta$. We will also need this inverse $\psi^{-1}(\eta) = \theta$ (available from the {\bf VGAM} package \citep{VGAM}) as well as its derivative $\partial \psi^{-1}(\eta)/ \partial \eta$, which we can estimate numerically.

    \[
    \psi(\lambda, x) = \left\{
    \begin{array}{lr}
    \{(x+1)^{\lambda}-1 \}/\lambda, & (x \ge 0, \lambda \ne 0)\\
    \log(x+1), & (x \ge 0, \lambda = 0)\\
    -\{(-x+1)^{2-\lambda}-1 \}/(2-\lambda), & (x < 0, \lambda \ne 0)\\
    -\log(-x+1), & (x < 0, \lambda = 0)
    \end{array}
    \right.
    \]

    In implementation, we estimate a unique hyperparameter $\lambda \in (-3,3)$ and apply a transformation for each of the model parameters independently.
    
    \item Approximate the model-based information $H_\eta \approx V^{-1}(\eta)$ as the inverse of the (pseudo-) posterior variance from the weighted sampler.
    \item Estimate the empirical information matrix
    $J^{\pi}_{\eta}(\eta) = J^{\pi}(\psi^{-1}(\eta)) * G(\eta)$
    where the multiplication $*$ is element-wise and the cells $\{G(\eta)\}_{j\ell} = (\partial \psi^{-1}(\eta_j)/ \partial \eta_j)(\partial \psi^{-1}(\eta_\ell)/ \partial \eta_\ell)$. We do this by first estimating $J^{\pi}(\hat{\theta})$ with $\bar{\theta} = \psi^{-1}(\bar{\eta})$ where $\bar{\eta}$ is the posterior mean. We then apply $G(\bar{\eta})$ due to the chain rule.
    \item Adjust the transformed parameters
    \[
\eta^{a}_m = \left(\eta_m - \bar{\eta}\right) R^{-1}_2 R_1  + \bar{\eta}.
\]
Then transform back to 
$\hat{\theta}^{a}_m = \psi^{-1}(\eta^{a}_m)$.
\end{itemize}

\subsection{Summary of Alternatives}
We now present the three approaches in the form of pseudo-code. Algorithm \ref{alg:naive} is an adaptation of the original approach from \cite{WilliamsISR21} where we (1) use a different replication approach (jackknife instead of bootstrap) and (2) are more explicit about the estimation of $H_\theta$ and $J^{\pi}_{\theta}$ using posterior gradients $\xi_{\theta}$ vs. likelihood gradients $\ell_{\theta}$.
Algorithm \ref{alg:curvature} modifies Algorithm \ref{alg:naive} by using an offset from the prior curvature and adding that to $J^{\pi}_{\theta}$ (changes in \textcolor{blue}{blue}).
Algorithm \ref{alg:YJ} modifies Algorithm \ref{alg:curvature} by including a variable transformation $\psi(\theta) = \eta$, applying adjustments to the transformed parameters, then back-transforming. In addition, Algorithm \ref{alg:YJ} replaces estimates of $H_\theta$ with the inverse of the empirical variance of the posterior draws $V^{-1}(\hat{\theta})$ (changes in \textcolor{blue}{blue}).

%try to modify algorithm example%
\IncMargin{1em}
\begin{algorithm}\label{alg:naive}
\DontPrintSemicolon
\SetKwInOut{Input}{input}\SetKwInOut{Output}{output}
\Input{
$\hat{\theta}_m$ from the pseudo-posterior \eqref{eq:pseudopost}\\
$\{j,k\}$ indicators for PSUs $j = 1,\ldots, J_{k}$ and Strata $k = 1,\ldots, K$. \\
$\{w_{ijk}, \mbf{X}_{ijk}\}$ for all $i$ in $1,\ldots, I_{jk}$ for every $\{j,k\}$\\
$R$ the total number of PSUs.
}
\Output{Adjusted sample $\hat{\theta}^{a}_m$}
\BlankLine
Calculate the posterior mean $\bar{\theta} = \frac{1}{M}\mathop{\sum}_{m=1}^{M}\hat{\theta}_m$\;
Calculate Monte Carlo average $\hat{H}_{\xi}  =\frac{-1}{M}\sum_{m=1}^{M}\mathop{\sum}_{i \in S} w_i \ddot{\xi}_{\theta_m}(\mbf{X}_{i})$\;
\For{\emph{PSUs} $r\leftarrow 1$ \KwTo $R$}{
Create replicate weights $w^{r}_{ijk} = w_{ijk}$\;
Set weight to zero for each member in $r^{th}$ PSU: $w^{r}_{irk} = 0$\;
Scale weights for remaining PSU's $\tilde{w}^{r}_{ir'k} = w_{ir'k} 
\left(\mathop{\sum}_{ij} w_{ijk}/ \mathop{\sum}_{ij} w^{r}_{ijk} \right)$\;
Evaluate $j_r = \mathop{\sum}_{ijk} \tilde{w}^{r}_{ijk} \dot{\xi}_{\bar{\theta}}(\mbf{X}_{ijk})$\; 
}
Calculate 	$\hat{J}^{\pi}_{\theta_{0}} = \frac{C}{R-1}\mathop{\sum}_{r=1}^{R}(j_r - \bar{j})(j_r - \bar{j})^t$ with $\bar{j} = \frac{1}{R}\mathop{\sum}_{r=1}^{R} j_r$\;
Calculate  $\hat{R}_1$ via Cholesky decomposition: $\hat{R}'_1 \hat{R}_1 = \hat{H}_{\xi}^{-1} \hat{J}^{\pi}_{\theta_{0}} \hat{H}_{\xi}^{-1}$\;
Calculate $\hat{R}_2$ via Cholesky decomposition: $\hat{R}'_2 \hat{R}_2 = \hat{H}_{\xi}^{-1}$\;
Calculate inverse $\hat{R}^{-1}_2$\;
Evaluate Eq. \ref{eq:adjustment}: $\hat{\theta}^{a}_m = \left(\hat{\theta}_m - \bar{\theta}\right) \hat{R}^{-1}_2 \hat{R}_1  + \bar{\theta}$\;
\caption{Na\"ive adjustment of pseudo-posterior}
\end{algorithm}\DecMargin{1em}

\IncMargin{1em}
\begin{algorithm}\label{alg:curvature}
\DontPrintSemicolon
\SetKwInOut{Input}{input}\SetKwInOut{Output}{output}
\Input{
$\hat{\theta}_m$ from the pseudo-posterior \eqref{eq:pseudopost}\\
$\{j,k\}$ indicators for PSUs $j = 1,\ldots, J_{k}$ and Strata $k = 1,\ldots, K$. \\
$\{w_{ijk}, \mbf{X}_{ijk}\}$ for all $i$ in $1,\ldots, I_{jk}$ for every $\{j,k\}$\\
$R$ the total number of PSUs.
}
\Output{Adjusted sample $\hat{\theta}^{a}_m$}
\BlankLine
Calculate the posterior mean $\bar{\theta} = \frac{1}{M}\mathop{\sum}_{m=1}^{M}\hat{\theta}_m$\;
\textcolor{blue}{Calculate the plug-in estimate for prior curvature $\hat{H}^{0}  = - \ddot{\log \pi}(\bar{\theta})$}\;
Calculate Monte Carlo average $\hat{H}_{\xi}  =\frac{-1}{M}\sum_{m=1}^{M}\mathop{\sum}_{i \in S} w_i \ddot{\xi}_{\theta_m}(\mbf{X}_{i})$\;
\For{\emph{PSUs} $r\leftarrow 1$ \KwTo $R$}{
Create replicate weights $w^{r}_{ijk} = w_{ijk}$\;
Set weight to zero for each member in $r^{th}$ PSU: $w^{r}_{irk} = 0$\;
Scale weights for remaining PSU's $\tilde{w}^{r}_{ir'k} = w_{ir'k} 
\left(\mathop{\sum}_{ij} w_{ijk}/ \mathop{\sum}_{ij} w^{r}_{ijk} \right)$\;
Evaluate $j_r = \mathop{\sum}_{ijk} \tilde{w}^{r}_{ijk} \dot{\xi}_{\bar{\theta}}(\mbf{X}_{ijk})$\; 
}
Calculate 	$\hat{J}^{\pi}_{\theta_{0}} = \frac{C}{R-1}\mathop{\sum}_{r=1}^{R}(j_r - \bar{j})(j_r - \bar{j})^t$ with $\bar{j} = \frac{1}{R}\mathop{\sum}_{r=1}^{R} j_r$\;
\textcolor{blue}{Adjust $\hat{J}^{\pi}_{\xi} = \hat{J}^{\pi}_{\theta_{0}} + \hat{H}^{0}$}\;
Calculate  $\hat{R}_1$ via Cholesky decomposition: $\hat{R}'_1 \hat{R}_1 = \hat{H}_{\xi}^{-1} \textcolor{blue}{\hat{J}^{\pi}_{\xi}} \hat{H}_{\xi}^{-1}$\;
Calculate $\hat{R}_2$ via Cholesky decomposition: $\hat{R}'_2 \hat{R}_2 = \hat{H}_{\xi}^{-1}$\;
Calculate inverse $\hat{R}^{-1}_2$\;
Evaluate Eq. \ref{eq:adjustment}: $\hat{\theta}^{a}_m = \left(\hat{\theta}_m - \bar{\theta}\right) \hat{R}^{-1}_2 \hat{R}_1  + \bar{\theta}$\;
\caption{Adjustment of pseudo-posterior using prior curvature (changes from Algorithm \ref{alg:naive} in \textcolor{blue}{blue})}
\end{algorithm}\DecMargin{1em}

\IncMargin{1em}
\begin{algorithm}\label{alg:YJ}
\DontPrintSemicolon
\SetKwInOut{Input}{input}\SetKwInOut{Output}{output}
\Input{
$\hat{\theta}_m = \{\hat{\theta}_1,\ldots,\hat{\theta}_D \}_m $ from the pseudo-posterior \eqref{eq:pseudopost}\\
$\{j,k\}$ indicators for PSUs $j = 1,\ldots, J_{k}$ and Strata $k = 1,\ldots, K$. \\
$\{w_{ijk}, \mbf{X}_{ijk}\}$ for all $i$ in $1,\ldots, I_{jk}$ for every $\{j,k\}$\\
$R$ the total number of PSUs.
}
\Output{Adjusted sample $\hat{\theta}^{a}_m$}
\BlankLine
\For{\emph{Parameters} $d\leftarrow 1$ \KwTo $D$}{
\textcolor{blue}{Estimate $\hat{\lambda}_d \in (-3,3)$} \;
\textcolor{blue}{Apply transformation $\{\eta_d\}_m = \psi(\hat{\lambda}_d, \{\hat{\theta}_d\}_m) $}\;
}
\textcolor{blue}{Compile $\eta_m =\left[\{\eta_1\}_m, \ldots, \{\eta_D\}_m \right]$ and $\psi(\theta) = 
\left[\psi(\hat{\lambda}_1, \{\hat{\theta}_d\}_m), \ldots \psi(\hat{\lambda}_1, \{\hat{\theta}_d\}_m) \right]$}\;
\textcolor{blue}{Calculate the posterior mean $\bar{\eta} = \frac{1}{M}\mathop{\sum}_{m=1}^{M}\eta_m$}\;
\textcolor{blue}{Estimate the posterior variance $V(\eta) = \frac{1}{M-1}\mathop{\sum}_{m=1}^{M} (\eta_m - \bar{\eta})(\eta_m - \bar{\eta})^{t}$}\;
\textcolor{blue}{Estimate the posterior curvature $\hat{H}_{\eta} = V^{-1}(\eta)$}\;
Calculate the plug-in estimate for prior curvature $\hat{H}^{0}  = - \ddot{\log \pi}(\textcolor{blue}{\psi^{-1}(\bar{\eta})})$\;
\For{\emph{PSUs} $r\leftarrow 1$ \KwTo $R$}{
Create replicate weights $w^{r}_{ijk} = w_{ijk}$\;
Set weight to zero for each member in $r^{th}$ PSU: $w^{r}_{irk} = 0$\;
Scale weights for remaining PSU's $\tilde{w}^{r}_{ir'k} = w_{ir'k} 
\left(\mathop{\sum}_{ij} w_{ijk}/ \mathop{\sum}_{ij} w^{r}_{ijk} \right)$\;
Evaluate $j_r = \mathop{\sum}_{ijk} \tilde{w}^{r}_{ijk} \dot{\xi}_{\psi^{-1}(\bar{\eta})}(\mbf{X}_{ijk})$\; 
}
Calculate 	$\hat{J}^{\pi}_{\theta} = \frac{C}{R-1}\mathop{\sum}_{r=1}^{R}(j_r - \bar{j})(j_r - \bar{j})^t$ with $\bar{j} = \frac{1}{R}\mathop{\sum}_{r=1}^{R} j_r$\;
Adjust on $\theta$ space $\hat{J}^{\pi}_{\xi} = \hat{J}^{\pi}_{\theta} + \hat{H}^{0}$\;
\textcolor{blue}{Calculate $G(\bar{\eta}) = \left(\frac{\partial \psi^{-1}(\eta)}{ \partial \eta}|_{\bar{\eta}}\right)
\left(\frac{\partial \psi^{-1}(\eta)}{ \partial \eta}|_{\bar{\eta}}\right)^{t}$}\;
\textcolor{blue}{Convert to $\eta$ space $\hat{J}^{\pi}_{\eta} = \hat{J}^{\pi}_{\xi} * G(\eta)$}\;
Calculate  $\hat{R}_1$ via Cholesky decomposition: $\hat{R}'_1 \hat{R}_1 = \hat{H}_{\eta}^{-1} \hat{J}^{\pi}_{\eta} \hat{H}_{\eta}^{-1}$\;
Calculate $\hat{R}_2$ via Cholesky decomposition: $\hat{R}'_2 \hat{R}_2 = \hat{H}_{\eta}^{-1}$\;
Calculate inverse $\hat{R}^{-1}_2$\;
Evaluate Eq. \ref{eq:adjustment}: $\eta^{a}_m = \left(\eta_m - \bar{\eta}\right) \hat{R}^{-1}_2 \hat{R}_1  + \bar{\eta}$\;
\textcolor{blue}{Transform back $\hat{\theta}^{a}_m = \psi^{-1}(\eta^{a}_m)$}\;
\caption{Adjustment of pseudo-posterior using transformations toward normality and prior curvature (changes from Algorithm \ref{alg:curvature} in \textcolor{blue}{blue})}
\end{algorithm}\DecMargin{1em}

%add Algorithm 3

%\section{Results}\label{sec:results}
\section{Simulations}\label{sec:sims}
We present three simulation studies to compare the unadjusted survey-weighted posterior to the three adjustment methods (e.g. na\"\i ve, prior-curvature, YJ transformation + prior-curvature). Our first study compares the methods for the case of a simple random sample, under which the unadjusted approach should be ideal and other methods should perform similarly in order to be seriously considered. Our second study induces a single-stage informative sampling design in which the data and group composition are skewed compared to the general population.  Our third study utilizes a two-stage cluster sample with both stages inducing an informative design. In these last two scenarios, we seek methods that improve upon the unadjusted pseudo-posterior. In general, these designs are `strongly' informative, in that both the distribution of responses $y_{ij}$ and the proportion of group membership $j$ in the sample are much different from that of the population. 
\subsection{SRS Simulation}
\begin{figure}
    \centering
    \includegraphics[width=.7\linewidth]{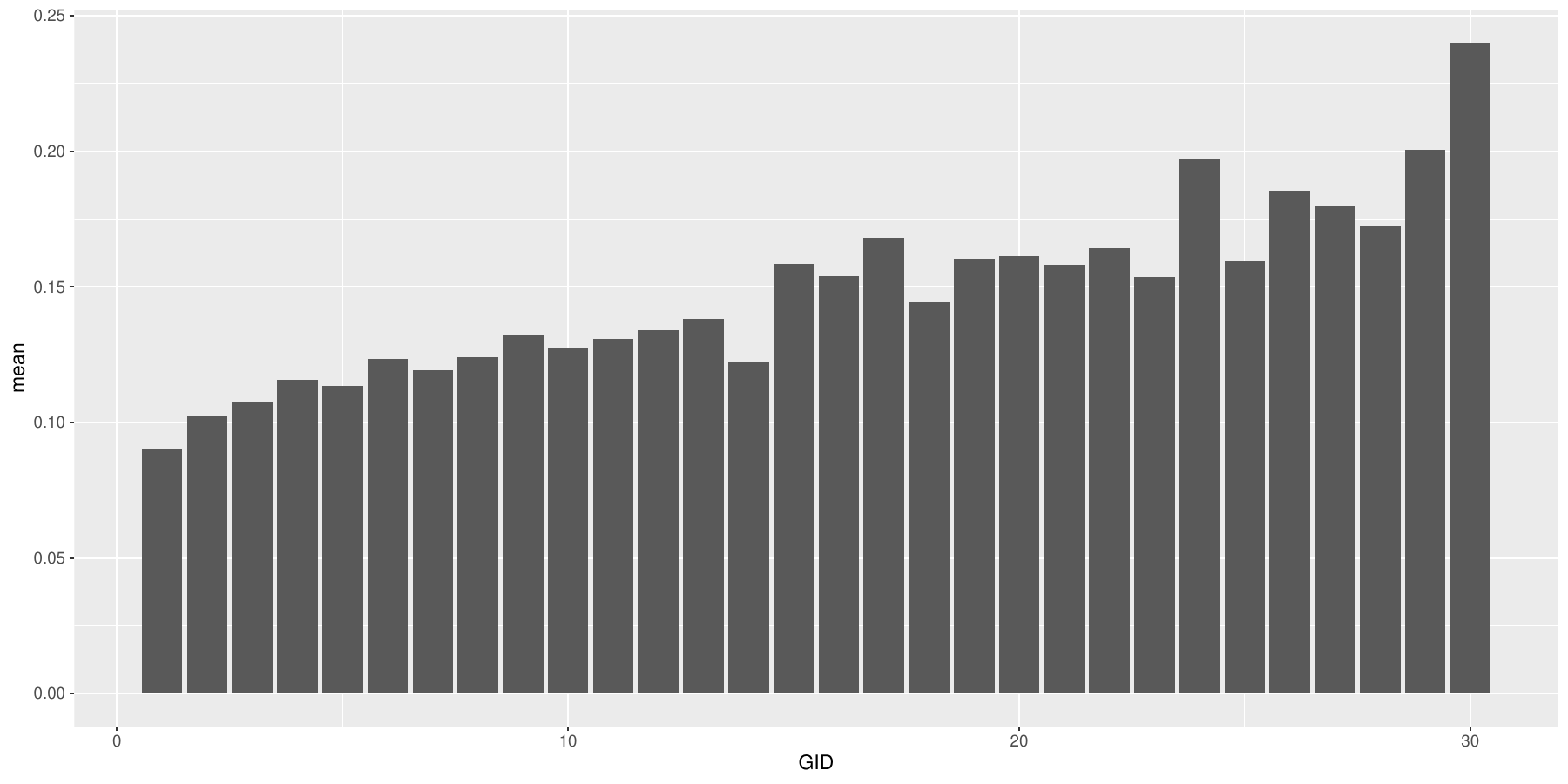}
    \caption{Proportion of binary outcomes in the simulated population across 30 groups.}
    \label{fig:depdist}
\end{figure}

We first generate a population of $N = $ 10,000 individuals, each assigned to one of $G= 30$ groups ranging in size $n_G$ from 2,000 to 100. We then generate binary outcomes from the following model:
\[
\begin{array}{rl}
    y_{ij} &\sim \ Bern(p_{ij})\\
    logit(p_{ij}) & = \  -2 + x_{ij} + \alpha_j\\
        \alpha_j & \sim \ N(0, 0.25)\\
    x_{ij} &\sim \ N(0,1)
\end{array}
\]
Figure \ref{fig:depdist} shows the distribution of the binary outcome across the 30 groups in the population. The proportion increases across groups, roughly doubling from the first to the last group.

We then take 200 simple random samples (SRS) sample of 1000 individuals. We estimate the survey-weighted pseudo-posterior on each and apply the different adjustments to the posterior variance. %\textcolor{violet}{add a plot of the proportion of y = 1 by group}
Table \ref{tab:SRS} compares the results over the 100 samples. As expected, the unadjusted pseudo-posterior distribution shows close to nominal coverage (95\%) on average for the fixed effects, random effects, and the random effects variance parameter. The three adjustments achieve or exceed coverage for fixed effects, incurring a cost in the increased size of the intervals. The na\"\i ve approach severely under-covers both the random effects and their variance. Both the prior curvature and the Yeo-Johnson approach (with prior curvature) dominate the na\"\i ve method in terms of coverage. Yeo-Johnson achieves nominal coverage on average for the random effects, but still falls somewhat short for the random effect variance. However, it gets much higher coverage than the na\"\i ve and prior curvature adjustments.

\begin{table}[h!]
    \centering
    \begin{tabular}{l|r|r|r|r|r|r|}
         & \multicolumn{3}{|c|}{Interval Length} & \multicolumn{3}{|c|}{Interval Coverage \%}  \\ \hline
    Adjustment     & Fixed & Random & $\sigma_\alpha$ & Fixed & Random & $\sigma_\alpha$ \\ \hline
    Unadjusted      &0.499	&1.277	&0.622	&93.0	&96.5	&96.0 \\
     Na\"\i ve          &0.532  &0.405  &0.146  &95.5   &52.7   &38.0 \\
Prior Curvature     &0.958	&0.980	&0.189	&99.5	&89.6	&46.0\\
     Yeo-Johnson    &0.566  &1.217  &0.377  &97.0   &95.6   &78.0\\ \hline
    \end{tabular}
    \caption{Results for 100 SRS samples, comparing mean interval length and coverage over fixed effects (slope and intercept), random effects (30 groups), and the random effect variance for the four alternative variance approaches.}
    \label{tab:SRS}
\end{table}

\subsection{PPS Simulation}
\begin{figure}
    \centering
    \includegraphics[width=.7\linewidth]{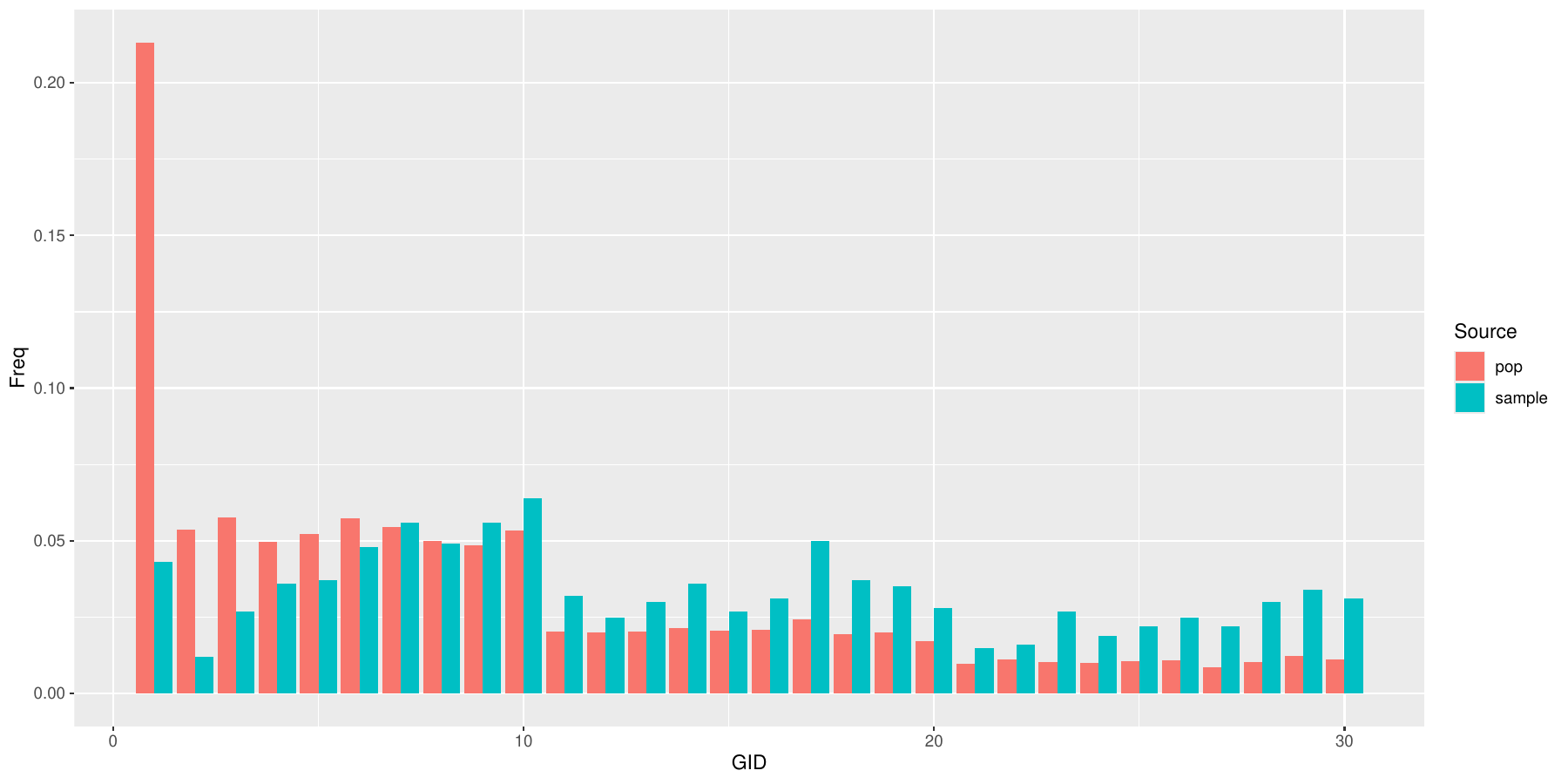}
    \caption{Comparison of proportion of group membership between the population and a PPS sample}
    \label{fig:ppsgrpdist}
\end{figure}

Similar to the SRS simulation, we first generate a population of $N = $10,000 individuals, each assigned to one of $G= 30$ groups ranging in size $n_G$ from 2,000 to 100. We use the same population generation model for $y_{ij}$. To create a Probability Proportional to Size (PPS) sample, we develop a size measure for each individual:

\[
s_{ij} = \max\{0.1, (\mu_{ij} - \min_{i,j}(\mu_{ij})) + 5 \alpha_j\}
\]
where $\mu_{ij} = logit(p_{ij})$. The size measure is larger for individuals more likely to have a value of 1 and is larger for those belonging to a group with a large random effect. Figure \ref{fig:ppsgrpdist} compares the distribution of group membership in the population with that from one PPS sample. Compared to the population, the PPS sample has a higher proportion of membership in groups with larger group IDs. These same groups also have a larger proportion of positive indicators (Figure \ref{fig:depdist}). Together this creates samples that have both a different proportion of positive responses and a larger representation of some groups compared to the population.% \textcolor{violet}{include a plot comparing the population distribution of groups to a sample. include a plot of the proportion of 1's across groups vs. pop and the overall sample}

Table \ref{tab:PPS} compares the results over the 100 repeated samples. As expected, the unadjusted pseudo-posterior distribution falls short of coverage due to the unequal weighting and the dependence induced by the sampling design. The na\"\i ve adjustment leads to some improvement in coverage for the fixed effects. It also has severe under-coverage for the random effects and the random effect variance. The Yeo-Johnson adjustment dominates both the unadjusted and na\"\i ve adjusted approaches, leading to some improvements for both the fixed effects and the random effects. The prior curvature approach has the highest coverage for fixed effects, but reduced coverage for the random effects and random effects variance compared to the unadjusted method. All approaches fall short of the nominal 95\% coverage level. This may be due to the relatively small sample size and strongly informative sample design.

\begin{table}[h!]
    \centering
    \begin{tabular}{l|r|r|r|r|r|r|}
         & \multicolumn{3}{|c|}{Interval Length} & \multicolumn{3}{|c|}{Interval Coverage \%}  \\ \hline
    Adjustment     & Fixed & Random & $\sigma_\alpha$ & Fixed & Random & $\sigma_\alpha$ \\ \hline
    Unadjusted      &0.531	&1.239	&0.557	&64.5	&86.8	&89.0 \\
     Na\"\i ve          &0.686  &0.377  &0.180  &71.0   &39.0   &39.0\\
Prior Curvature     &1.111	&0.981	&0.232	&88.5	&79.1	&51.0\\
     Yeo-Johnson    &0.658  &1.677  &0.766  &72.0   &88.9   &90.0\\ \hline
    \end{tabular}
    \caption{Results for 100 PPS samples, comparing mean interval length and coverage over fixed effects (slope and intercept), random effects (30 groups), and the random effect variance for the four alternative variance approaches.}
    \label{tab:PPS}
\end{table}

\subsection{Two-Stage Sample Simulation} We first generate a larger population of $N = $100,000 individuals, each assigned to one of $G= 30$ groups ranging in size $n_G$ from 20,000 to 1,000. In order to create an informative two-stage design, we first sort the individuals by their true expected value $p_{ij}$. We then assign 1000 primary sampling units using the sort order: the largest 100 get assigned to the first PSU, the next largest 100 get assigned to the second PSU and so on. Thus individuals within each PSU are more similar to each other than to a randomly selected member of the population. We develop a size measure for each PSU:
\[
s_{k} = \max\{0.1, (\mu_{k} - \min_{k}(\mu_{k}))^2 \}
\]
where $\mu_{k} = \frac{1}{100} \sum_{i,j} \mu_{ij} 1_{k}(i,j)$ is the average over the 100 individuals in the $k^{th}$ PSU. We then select 100 of the 1,000 clusters (PSUs) from the population.

For the second stage, within each PSU indexed by $k$, we select individuals with probability proportional to size:
\[
s^{k}_{ij} = \max\{0.1, (\mu_{ij} - \min_{i,j}(\mu_{ij})) + 5 (\alpha_j - \min_{j}(\alpha_j))\}
\]
We select 10 individuals from each cluster for a total sample size of 1,000.

\begin{figure}
    \centering
    \includegraphics[width=.7\linewidth]{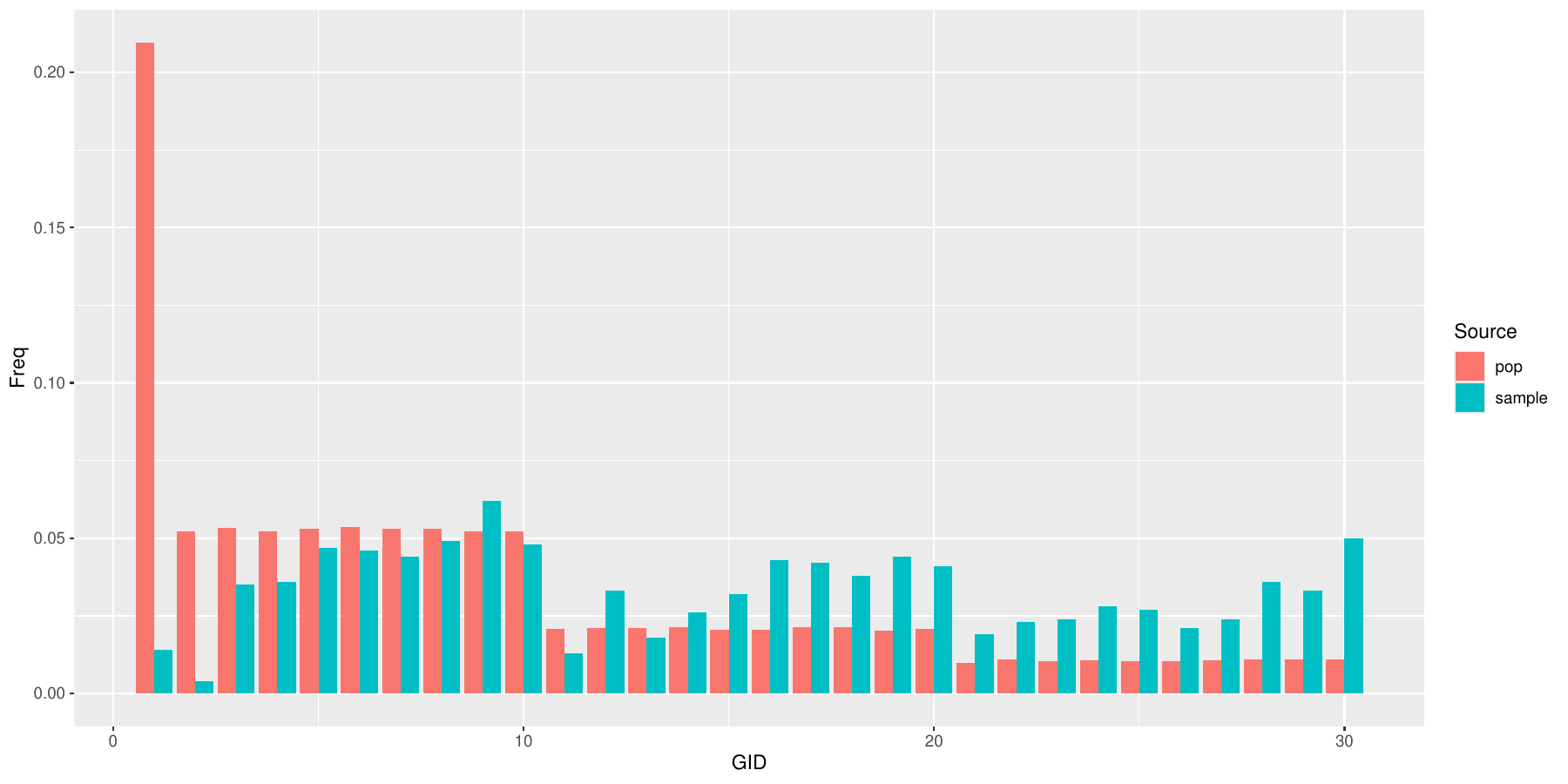}
    \caption{Comparison of proportion of group membership between the population and a multistage PPS sample}
    \label{fig:multigrpdist}
\end{figure}

At the first stage, the size measure is larger for clusters that have a higher proportion of 1's as well as a higher proportion of groups with large random effects. Together this should induce a design that has both a different proportion of positive responses and a larger representation of some groups compared to the population. The second stage of sampling amplifies this imbalance by further selecting individuals within each cluster which are more likely to have a response of 1 and more likely to be from a group with a larger random effect. This can be seen in Figure \ref{fig:multigrpdist} in which the group membership in the multistage sample is further skewed towards the right (the rarer groups with larger proportions of the binary response). %\textcolor{violet}{include a plot comparing the population distribution of groups to a sample. include a plot of the proportion of 1's across groups vs. pop and the overall sample}

Table \ref{tab:TSS} compares the results over the 100 repeated samples. The unadjusted pseudo-posterior distribution falls short of coverage for the fixed effects parameters and the random effect variance. However, it seems to cover the random effects at a close to nominal rate.  The na\"\i ve adjustment leads to almost no improvement for the fixed effects and much worse coverage for the random effects and the random effects variance. The Yeo-Johnson adjustment dominates both the unadjusted and na\"\i ve adjusted approaches, leading to modest improvements for the fixed effects and the random effects. Coverage for the random effects variance is much improved, but still far from nominal. The prior curvature approach improves coverage for fixed effects and random effects but shows little improvement in coverage for the random effects variance compared to the na\"\i ve approach. All approaches fall short of the nominal 95\% coverage level for the fixed effect and the random effects variance. This may be due to the relatively small sample size and strongly informative sample design.

\begin{table}[h!]
    \centering
    \begin{tabular}{l|r|r|r|r|r|r|}
         & \multicolumn{3}{|c|}{Interval Length} & \multicolumn{3}{|c|}{Interval Coverage \%}  \\ \hline
    Adjustment     & Fixed & Random & $\sigma_\alpha$ & Fixed & Random & $\sigma_\alpha$ \\ \hline
    Unadjusted      &0.643	&1.896	&0.678	&62.5	&92.2	&21.0 \\
     Na\"\i ve      &0.662	&1.277	&0.417	&65.0	&76.6	&8.0\\ 
Prior Curvature     &0.883	&1.908	&0.459	&84.5	&93.6	&9.0\\
     Yeo-Johnson    &0.744	&2.538	&1.091	&72.0	&94.6	&40.0\\ \hline
    \end{tabular}
    \caption{Results for 100 two-stage samples, comparing mean interval length and coverage over fixed effects (slope and intercept), random effects (30 groups), and the random effect variance for the four alternative variance approaches.}
    \label{tab:TSS}
\end{table}

\section{NSDUH Application}\label{sec:NSDUH}
Our motivating example comes from the analysis of depression in adults from the National Survey on Drug Use and Health (NSDUH) by \cite{10.1093/aje/kwae121}. The NSDUH is a nationally representative survey of the  civilian, non-institutionalized population aged 12 years or older residing within the 
United States. The multi-stage design stratifies states and regions within states, then takes a series of nested samples of increasing geographic resolution leading to the selection of individual dwelling units and up to two individuals within each. The survey over-samples smaller population states and younger age groups \citep{CBHSQ2020}. National level analyses across large age-ranges (such as all adults) will need to use the survey weights to mitigate bias due to differences in substance use and mental health rates across age groups and across states. The clustering of the sample by geographic areas (census block groups) and the sampling of up to two persons per household reduces the effective sample size of the design relative to a simple random sample. Together, the unequal selection probability and the clustering are non-ignorable features of the survey design. They should be included during analysis to mitigate bias and to obtain more accurate uncertainty quantification.

The full example uses the public use files from 2015 through 2020, resulting in a sample size of about 236,000 adults. There are 42 unique groups defined by the intersection of race/ethnicity, gender, and sexual orientation. For analysis, we use past year depression status and a multi-level logistic regression \eqref{eq:logisticMM} with fixed effects as the main effects (race/ethnicity, gender, sexual orientation) and the random effects as the intersectional groups. Our focus of estimation and uncertainty adjustment will be on the model parameters. However, this will also impact derived quantities such as group-level estimates. In the full model of \cite{10.1093/aje/kwae121}, they also include age as a fixed effect to use in age-adjusted estimates. For ease of discussion we exclude age from our example analysis.

We compare the unadjusted estimates (directly from pseudo-posterior) with the Yeo-Johnson (with prior curvature) adjustment approach. We compare results with fitting a posterior for each of the 100 replicate data sets and then taking the variance of the posterior means across replicates. Because this is computationally intensive (roughly 100 times slower), we first randomly subset the data. We subsample 500 adults from each of the 42 groups (taking all members of groups with fewer than 500 individuals). We then adjust the weights of each group to account for the sub-sampling. This leads to a sample of approximately 14,000 adults while still preserving the sample size of the smallest groups (For a comparison of the unadjusted and YJ adjusted estimates for this subsample, see Appendix \ref{app:subsample}. See Appendix \ref{app:runtime} for approximate run times).

\begin{figure}[h]
    \centering
    \includegraphics[width=1\linewidth, page = 2]{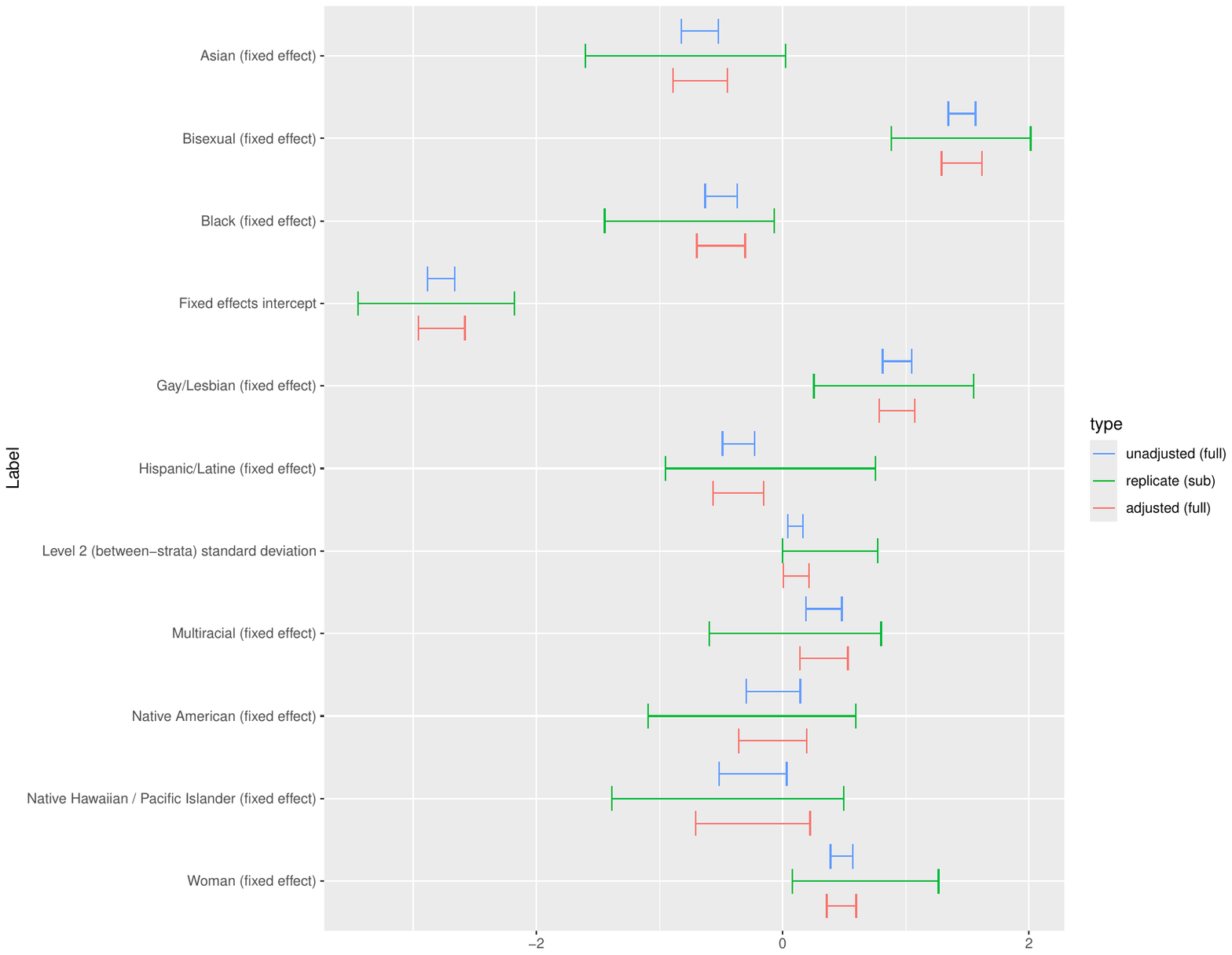}
    \caption{Comparison of 95\% intervals for global regression parameters for the multi-level logistic regression for past year depression from NSDUH.}
    \label{fig:NSDUHglobalfull}
\end{figure}

Figure \ref{fig:NSDUHglobalfull} compares the three alternative approaches in terms of their 95\% intervals for global parameters (fixed effects and random effect variance). The unadjusted sample has narrower intervals that the YJ adjusted sampled. This is to be expected, as typically we expect a `design effect' to reduce the effective sample size due to clustered and unequal sampling. Of note, the interval for random effect variance (Level 2) is narrowest for the unadjusted sample. We further note that the replication based approach results in larger intervals across all these global parameters due to the reduced size of the subsample (14,000 vs. 236,000).

\begin{figure}[h]
    \centering
    \includegraphics[width=1\linewidth, page = 3]{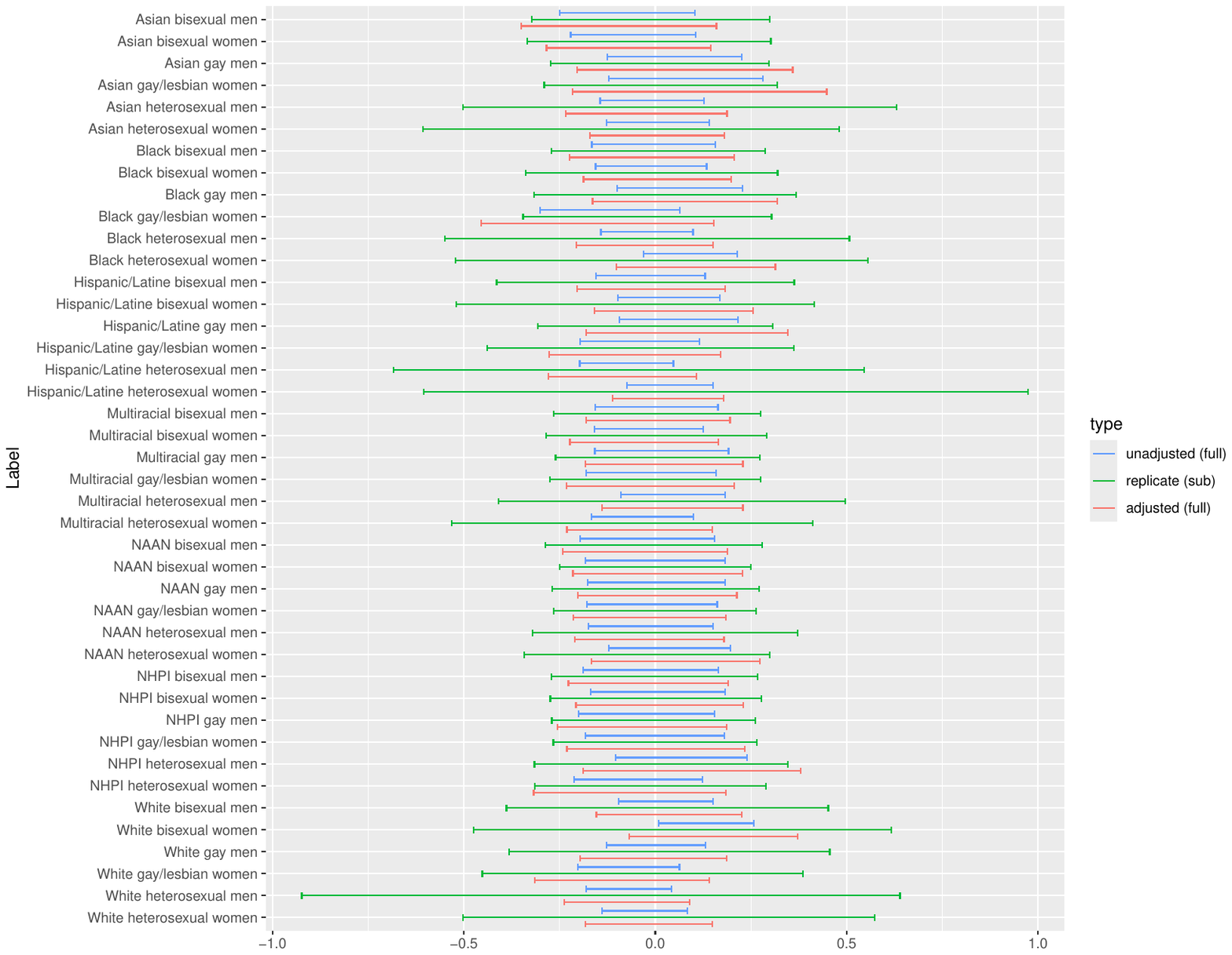}
    \caption{Comparison of 95\% intervals for local random effects for the multi-level logistic regression for past year depression from NSDUH.}
    \label{fig:NSDUHlocalfull}
\end{figure}

We next compare the performance of the three alternatives for random effects estimation (Figure \ref{fig:NSDUHlocalfull}). We see that the unadjusted intervals are the narrowest. The YJ intervals are consistently wider than those of the unadjusted approach. The naive replication method (using a subsample) has wider intervals for some random effects corresponding to larger groups, but similar intervals for smaller groups with original sample sizes less than 500, for example Native Hawaiian and Pacific Islander (NHPI) and Native American and Alaskan Native (NAAN) groups.

\begin{figure}[h]
    \centering
    \includegraphics[width=1\linewidth]{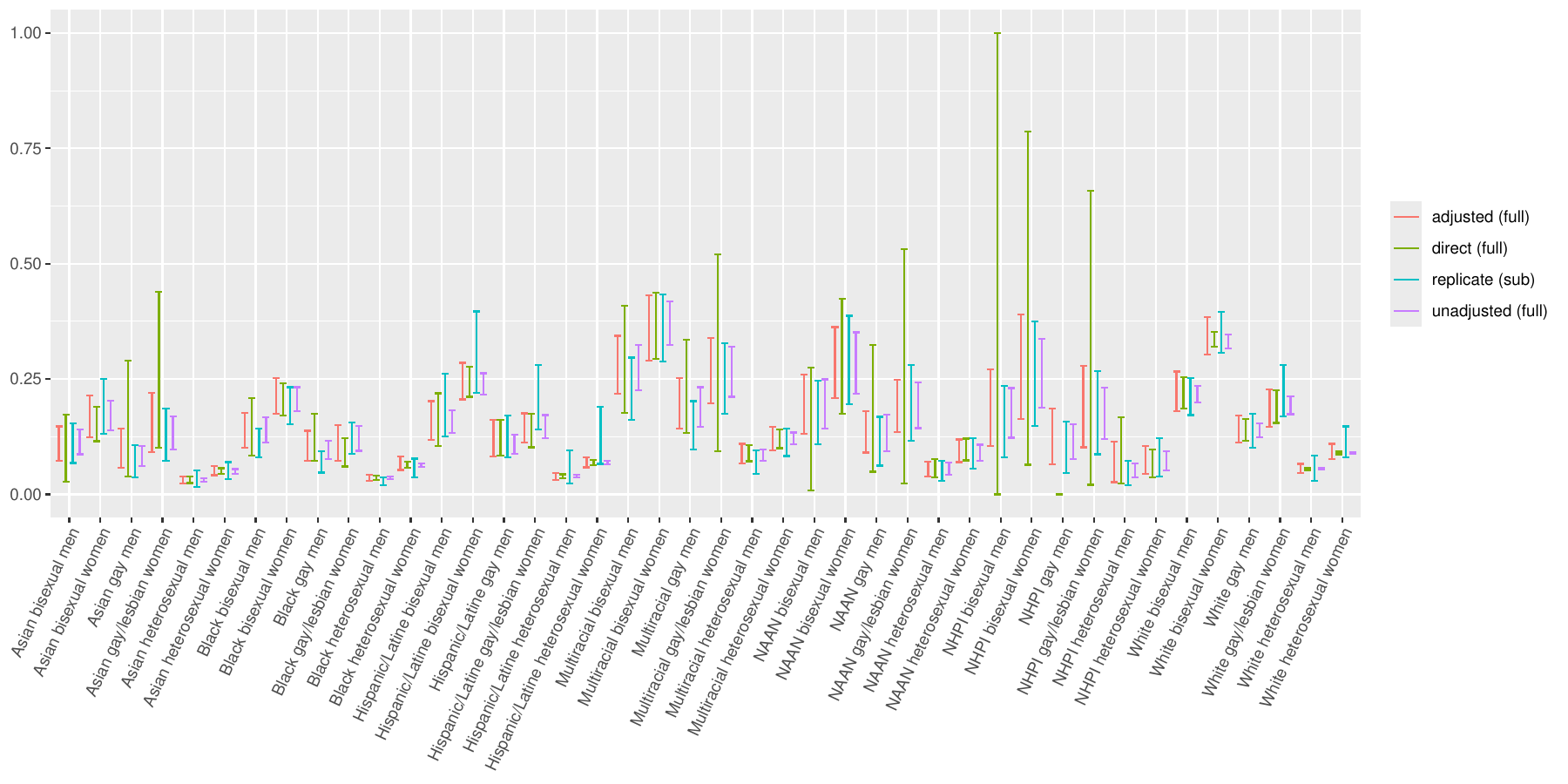}
    \caption{Comparison of 95\% intervals for group level estimates for past year depression from NSDUH.}
    \label{fig:NSDUHgroupfull}
\end{figure}

 Figure \ref{fig:NSDUHgroupfull} compares the resulting domain level estimates for each group to direct estimates from the full sample obtained by taking the weighted average depression rate within each group. In general, we see that for smaller sized groups (such as NHPI and NAAN) the direct estimates have extremely wide intervals compared to the model-based approaches. This is the main motivation for using models to gain precision for smaller domains. In contrast, for groups with larger sample sizes (such as White) the direct estimates have intervals of similar or even smaller widths compared to the model-based estimates. As expected, the YJ intervals are wider than the unadjusted intervals. Similar to the comparison of the random effects, the naive replication approach on the subsample leads to similar results to the YJ approach for smaller sized groups, but leads to wider intervals for larger and mid-sized groups (such as Hispanic and White).

\section{Discussion and Conclusions}\label{sec:discuss}
In this work, we demonstrated how an approach for automated adjustment of Bayesian models for survey data - which works well for global or fixed effects models - has challenges for fitting multi-level models with local parameters. Our motivating example was a variance component model for a logistic regression with random intercepts for different subpopulations. We discussed how the automation approach could be modified to be less reliant on asymptotics (e.g. large sample sizes) through the incorporation of prior curvature and through automated transformations of the parameter space towards symmetry and normality. Through a series of simulations, we see that these modifications mitigate the underestimation of variance from the existing approach. Our application to past year depression from the NSDUH compares our best approach (YJ) to the brute force replication approach in which the Bayesian model is re-estimated for each replicate (e.g. 100 times) but on a stratified subsample of the data. Given a similar computational level of effort, our proposed approach on the full sample (YJ) provides more efficient intervals compared to a replication method applied to a subsample of the data. Both approaches provide improvements over direct estimates for small and mid-sized domains.
%REMOVE We could also consider  simulations incorporating the naive replication approach to confirm whether it will achieve appropriate coverage for the random effects.
%Alternatives for transformations - iterate or use more complext

Our motivating ANOVA-like model is a `shallow' hierarchical model. We also evaluated our approach on a `deeper' hierarchical selection model (global-local). Appendix \ref{app:globallocal} investigates a more complex model with a deeper hierarchy using global-local priors for a subset of NSDUH data. While the pseudo-likelihood approach leads to similar group-level point estimates, the YJ adjustment is extremely unstable for the more complex model. While the unadjusted estimates from the pseudo-posterior lead to similar domain level estimates between the simpler and more complex models, the YJ adjustment is very unstable and leads to unusable intervals. 

Future work could incorporate refinements to the YJ transformation. For example, we may treat the YJ transformed variables as skew-normal and provide one additional transformation towards normality before creating the sandwich adjustment (See \cite{robbins2013imputation} for an example used in imputation).
Future work could also focus on diagnosing when a model is `too deep' or when parameters are close to mixed type (point mass at zero) such that the resulting sandwich adjustment (even after transformation) is not appropriate. The `automatic' adjustment could be applied differentially, adjusting only a subset of parameters. This could be informed by diagnostics, such as data cloning \citep{lele2010estimability}, which attempt to estimate the rate of decrease of the variance to identify parameters that are not estimable. We could re-purpose this diagnostic to identify which parameters have variance reduction with increasing sample size in the global ($n^{-1}$), local ($n^{-1/2}$), or `deep' hyper-parameter (e.g. $1/\log(n)$) regimes.

%% Appendices %%
%%%%%%%%%%%%%%%%%%%%%%
\begin{appendix}
\section{NSDUH subsample comparisons}\label{app:subsample}
\begin{figure}[h]
    \centering
    \includegraphics[width=1\linewidth, page = 2]{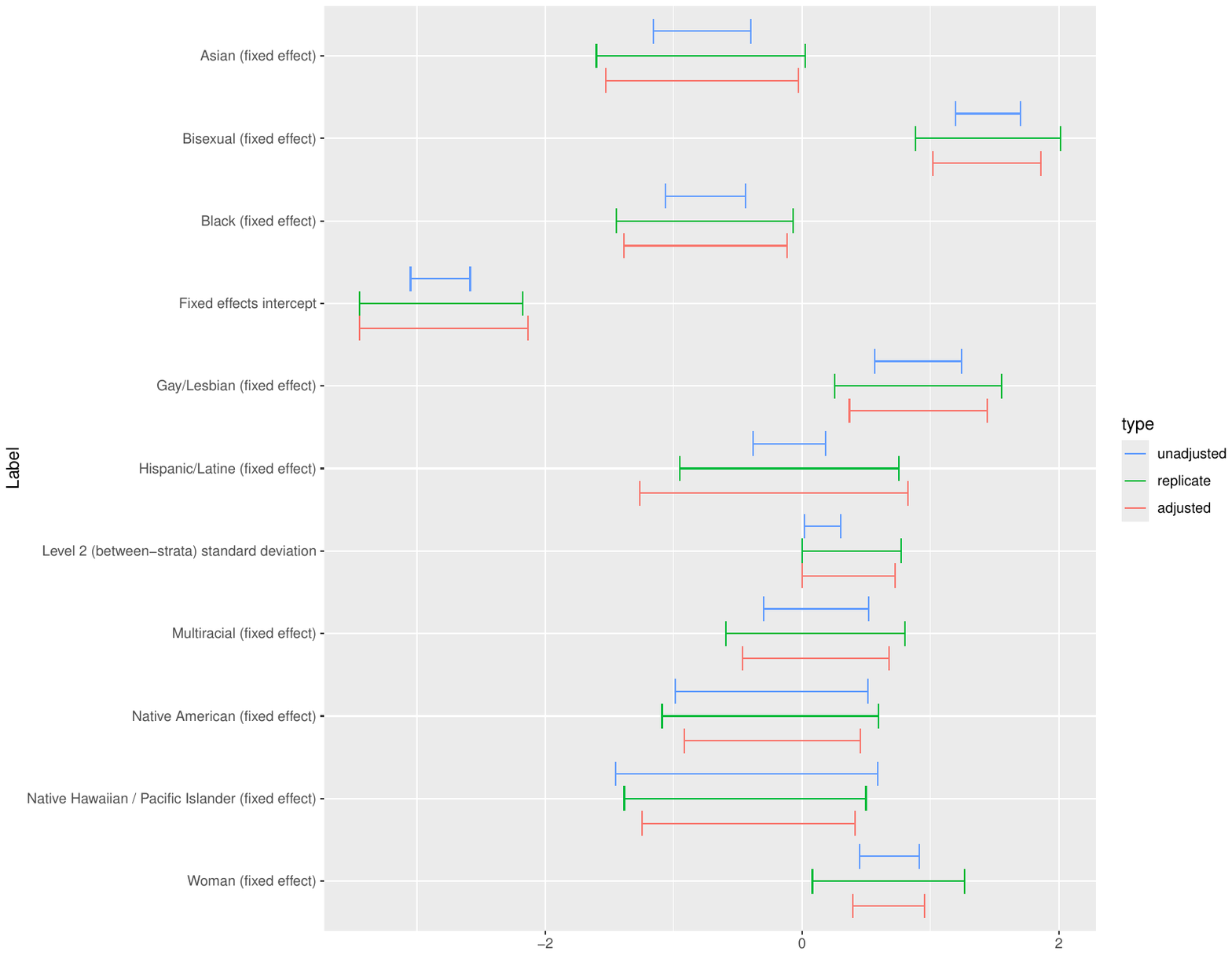}
    \caption{Comparison of 95\% intervals for global regression parameters for the multi-level logistic regression for past year depression from NSDUH.}
    \label{fig:NSDUHglobal}
\end{figure}

Using just the subsample of 14,000, 
we compare the unadjusted estimates (direct from pseudo-posterior) with those from the Yeo-Johnson adjustment approach and with estimates obtained by fitting a new Stan model for each of the 100 replicates. The latter is roughly 100 times slower but provides a reasonable alternative comparison. Figure \ref{fig:NSDUHglobal} compares the three alternatives in terms of their 95\% intervals for global parameters (fixed effects and random effect variance). The unadjusted posterior consistently has narrower intervals that either the YJ adjusted or the naive replication approach. This is to be expected, because a `design effect' reduces the effective sample size due to clustered and unequal sampling. Of note, the credible interval for the random effect variance (Level 2) is significantly narrower for the unadjusted method, while the two alternative approaches show similar results. We note that for the fixed effect for sex (women), the YJ adjusted interval is notably smaller than the replication interval but still slightly larger than the unadjusted interval.

\begin{figure}[h]
    \centering
    \includegraphics[width=1\linewidth, page = 3]{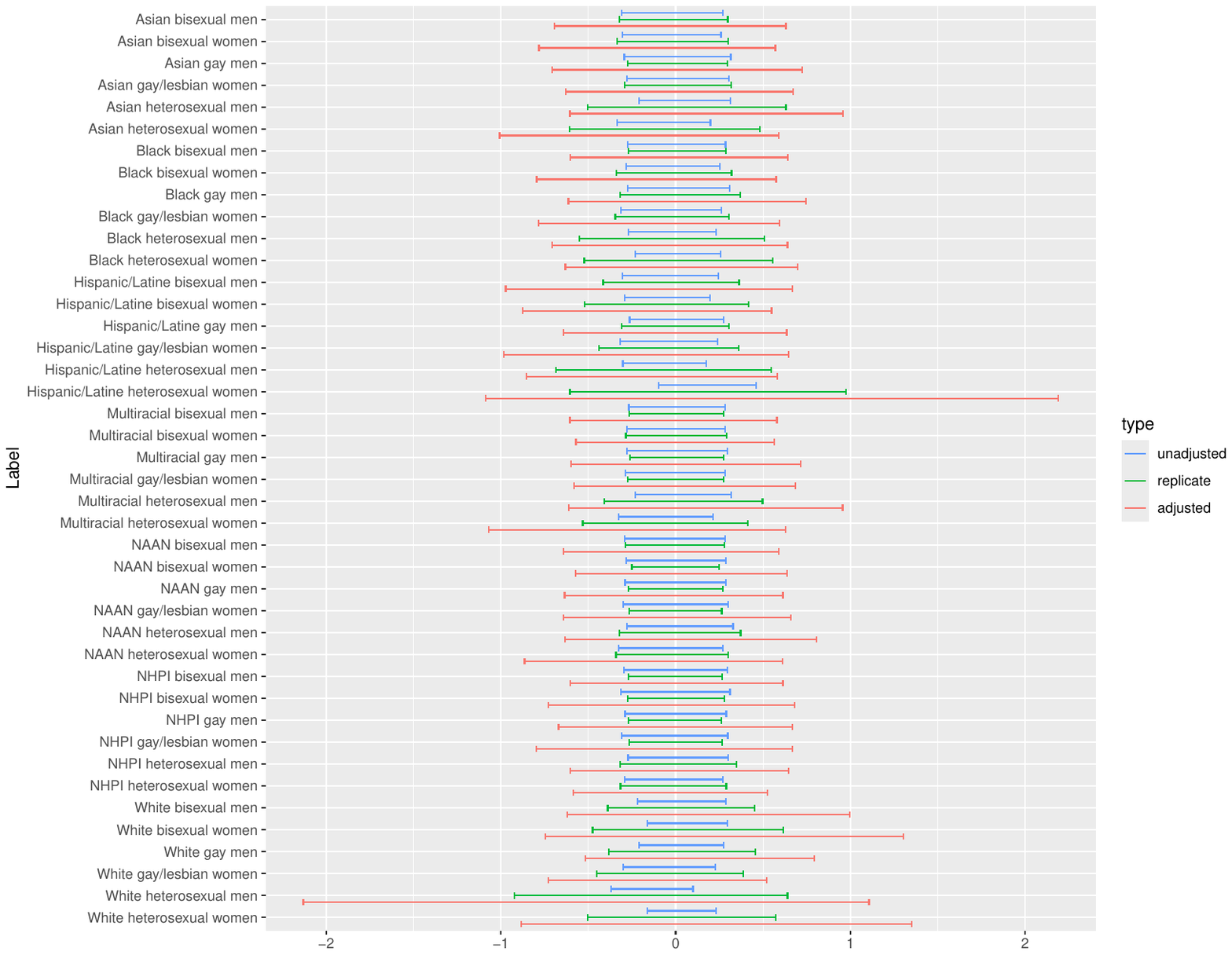}
    \caption{Comparison of 95\% intervals for local random effects for the multi-level logistic regression for past year depression from NSDUH.}
    \label{fig:NSDUHlocal}
\end{figure}

\begin{figure}[h]
    \centering
    \includegraphics[width=1\linewidth]{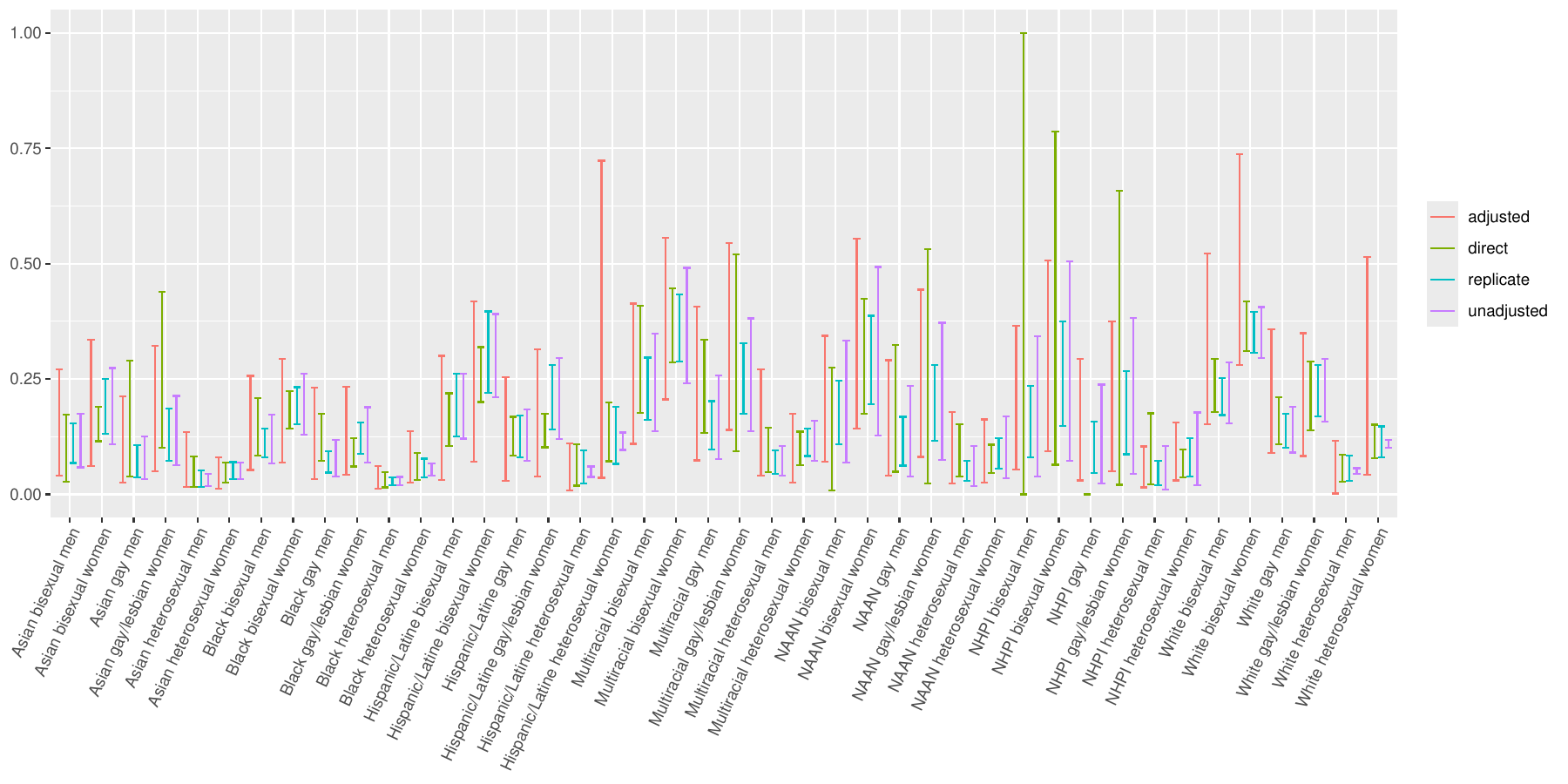}
    \caption{Comparison of 95\% intervals for strata/group level estimates for past year depression from NSDUH.}
    \label{fig:NSDUHgroup}
\end{figure}

We next compare the three approaches for local parameters or random effects (Figure \ref{fig:NSDUHlocal}. We see that the unadjusted intervals are the narrowest, with the naive replication approach leading to wider intervals. The YJ intervals are consistently wider than those for the replication approach. This is in contrast to the global parameters in which both intervals behaved similarly, including for the random effects (Level 2) variance. The wider intervals from the YJ adjustment translate into wider intervals for the domain estimates. Figure \ref{fig:NSDUHgroup} compares the three model-based alternatives to the direct estimates.

%Update this figure with group means and design based estimates

In contrast to the reasonable performance of the YJ intervals for the full sample, the instability of the YJ intervals for the subsample may be due to the extreme weights from the down-sampling of the larger groups. This indicates that the YJ adjustment for random effects may be sensitive to the scaling and stability of weights.

\section{Global-Local Priors for Interactions}\label{app:globallocal}
Our motivating hierarchical model \eqref{eq:logisticMM} is relatively simple or `shallow'. Instead of placing a simple prior on the group level random effects $\alpha_j$ defined by cross-classification of three variables we could instead follow \cite{si2020bayesian} and create a hierarchical prior structure that penalizes main effects as well as second-order and third-order interaction terms. These priors are linked, with second-order and third-order priors being amplified by priors from corresponding main effects. The resulting structure induces sparsity, particularly for higher-level interaction terms.

\begin{equation}\label{eq:logisticgloballocal}
\begin{array}{rl}
    y_{ij} &\sim \ Bern(p_{ij})\\
    logit(p_{ij}) & = \ \mu_{ij} = X_{ij} \beta + \alpha_j\\
    \beta & \sim \ N(0, \Sigma)\\
    %\Sigma & \sim \ G\\
    \alpha_j & = \sum_{k\in S^{(1)}} \alpha^{(1)}_{j,k} + \sum_{k\in S^{(2)}} \alpha^{(2)}_{j,k} + \sum_{k\in S^{(3)}} \alpha^{(3)}_{j,k}
\end{array}
\end{equation}
where $S^{(\ell)}$ is all $\ell-$way interactions and $\alpha^{(\ell)}_{j,k}$ is the $k^{th}$ interaction term of order $\ell$ that corresponds to group $j$. Following \cite{si2020bayesian}, the prior structure is
\begin{equation}\label{eq:logisticgloballocalpriors}
\begin{array}{rl}
    \alpha^{(\ell)}_{j,k} & \sim N\left(0, (\lambda_{k}^{(\ell)} \sigma)^2\right)\\
    \sigma & \sim \text{Cauchy}_{+}(0,1)\\
    \lambda_{k}^{(\ell)} & = \delta^{(\ell)} \prod_{\nu \in M^{(k)}} \lambda_{\nu}^{(\ell)}\\
    \lambda_{k}^{(1)} & \sim N_{+}(0,1)\\
    \delta^{(\ell)} & \sim N_{+}(0,1)
\end{array}
\end{equation}
where $\sigma$ is a `global' scale and each $\lambda_{k}^{(\ell)}$ is a local scale adjustment, such that higher order local scales depend on the scales from corresponding main effect terms from the set $M^{(k)}$. Lastly, $\delta^{(\ell)}$ is an additional scaling specific to the order of the interaction term.

\begin{figure}
    \centering
    \includegraphics[width=1\linewidth, page = 2]{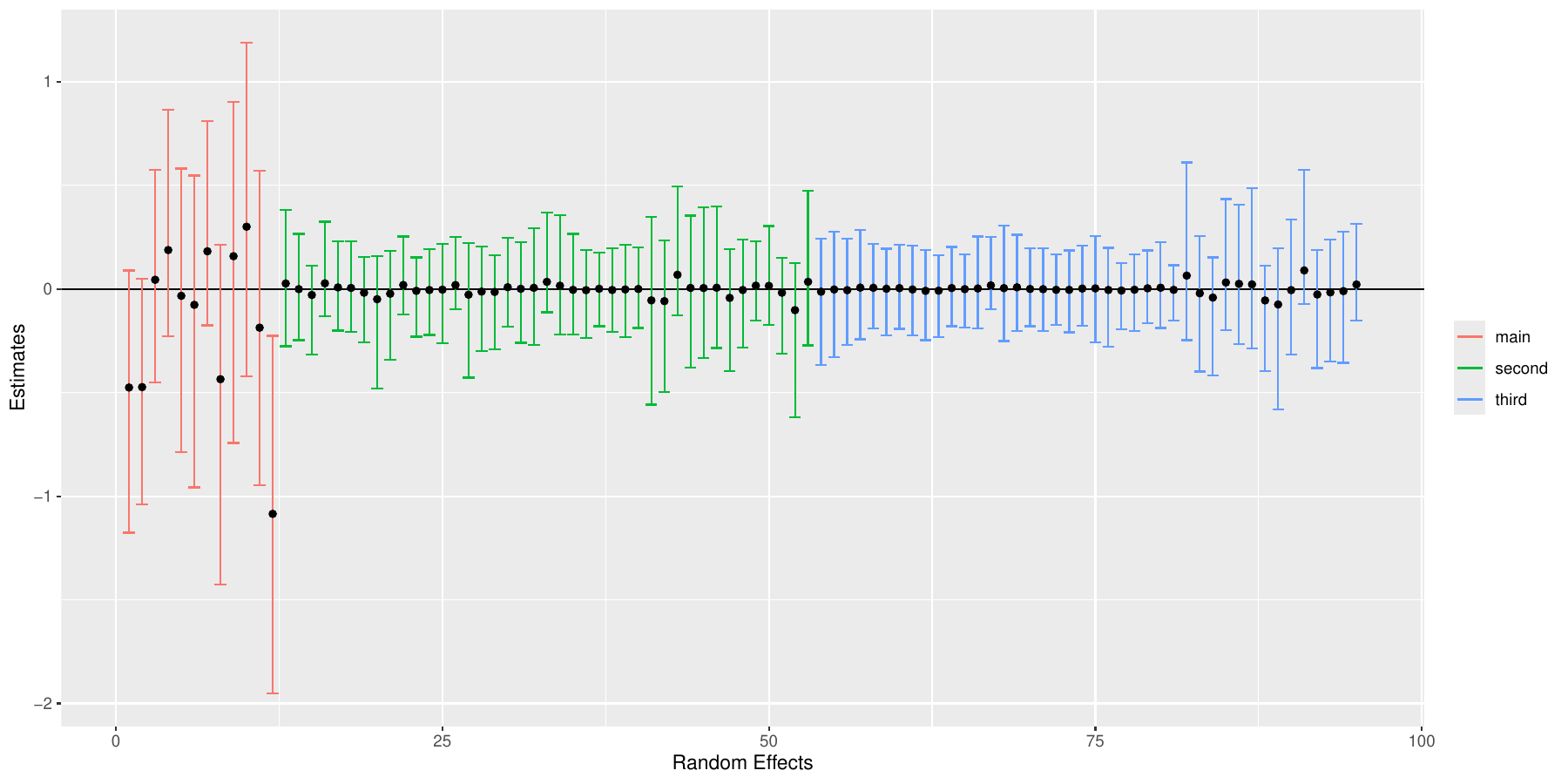}
    \caption{(Pseudo-)Posterior distribution (mean and 95\% interval) of random effects from the hierarchical global-local model estimated on a stratified subsample from NSDUH.}
    \label{fig:globallocalre}
\end{figure}

\begin{figure}
    \centering
    \includegraphics[width=1\linewidth, page = 2]{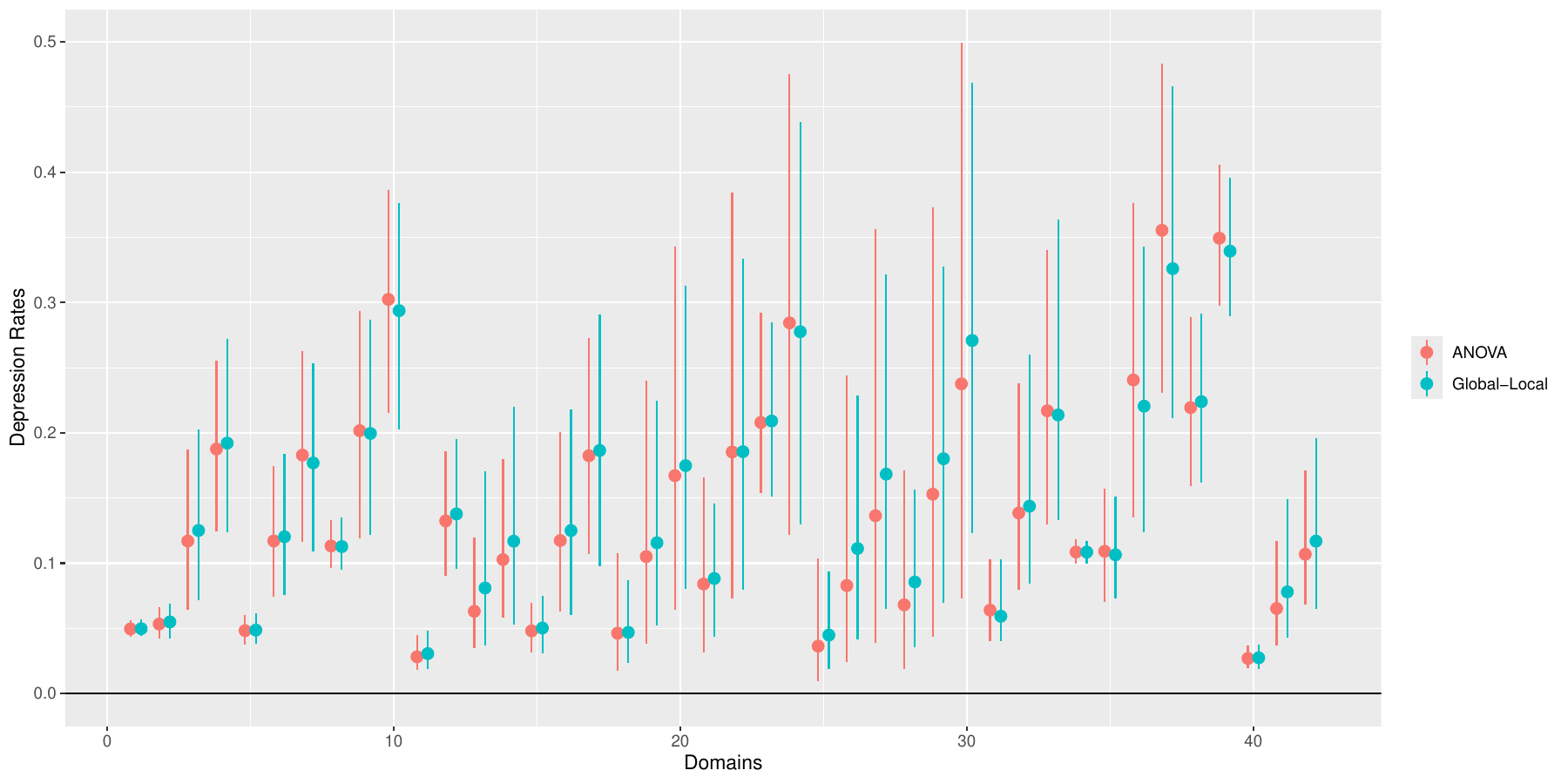}
    \caption{Comparison of domain level estimates of depression (mean and 95\% interval) from the \emph{unadjusted} survey-weighted ANOVA model and the hierarchical global-local model for a stratified subsample from NSDUH.}
    \label{fig:globallocalun}
\end{figure}

\begin{figure}
    \centering
    \includegraphics[width=1\linewidth, page = 2]{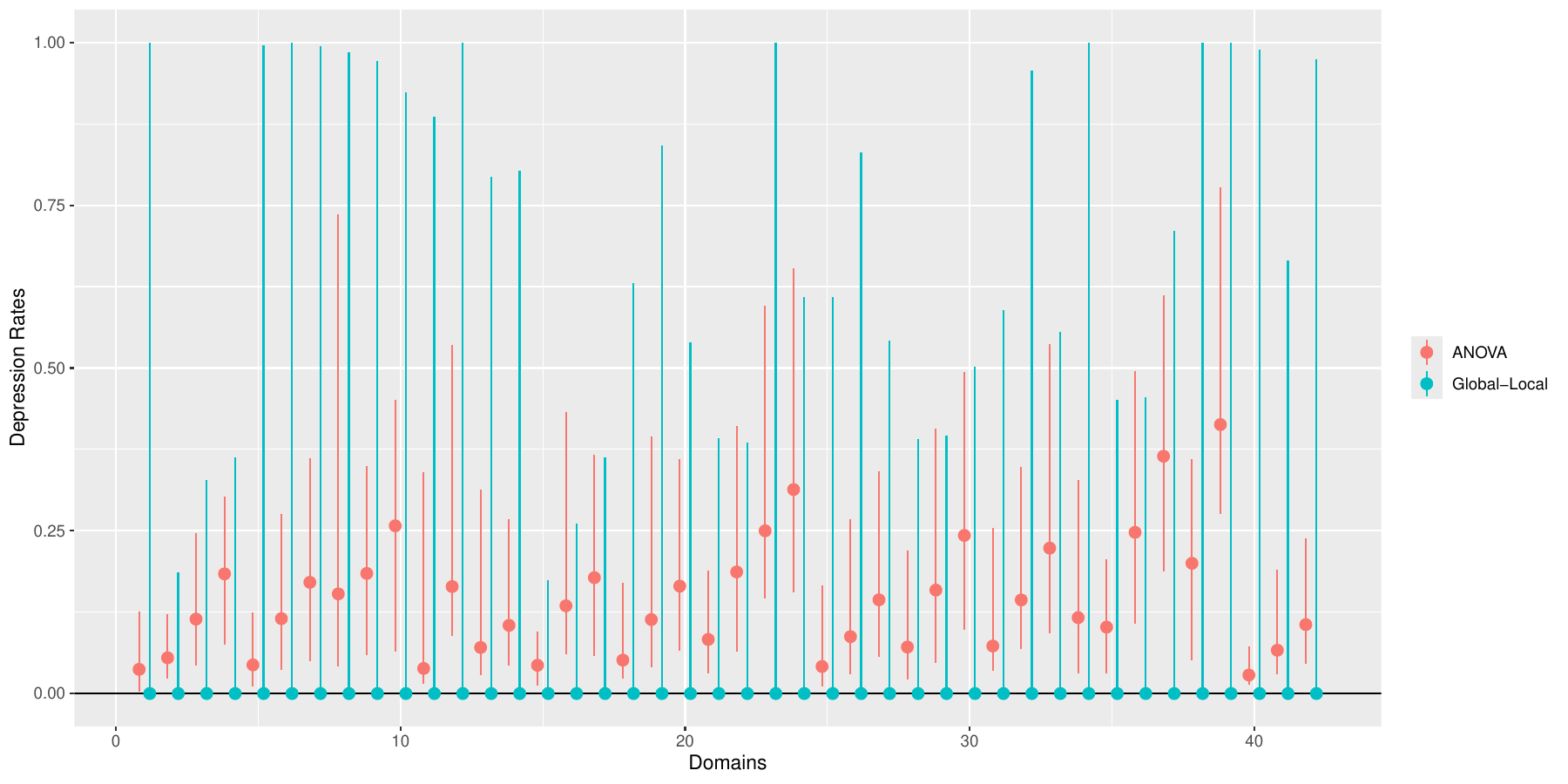}
    \caption{Comparison of domain level estimates of depression (mean and 95\% interval) from the \emph{adjusted} survey-weighted ANOVA model and the hierarchical global-local model for a stratified subsample from NSDUH.}
    \label{fig:globallocaladj}
\end{figure}

For demonstration, we estimate this more complex hierarchical model on the NSDUH stratified subsample of 14,000. Figure \ref{fig:globallocalre} shows (pseudo-) posterior mean and 95\% intervals for the random effects, color-coded by order (main, second, third). We see that the structured prior pulls most higher order random effects towards zero. However some second-level and third-level interactions are still noticeable. Comparing the adjusted (pseudo-) posterior estimates for domains between our simpler model \eqref{eq:logisticMM} and the more complex global-local \eqref{eq:logisticgloballocal}, we see qualitatively very similar estimates for domains (Figure \ref{fig:globallocalun}). 
From Appendix \ref{app:subsample}, we know that the YJ adjustment is unstable for the extreme subsampling for this NSDUH subset. The adjusted version of the global-local model is extremely unstable, leading to unusable results (Figure \ref{fig:globallocaladj}). This suggests a real limit to the `automatic' adjustment for deep hierarchical models. However, future work could investigate a tiered approach in which some hierarchical parameters are excluded from the adjustment.

\section{Run Time Comparisons}\label{app:runtime}
In this section, we summarize approximate run times for the different examples and alternative approaches. All runs were performed on the same personal computer using a single core per run. Table \ref{tab:runtime} summarizes the approximate run times. For example, the simulation studies of Section \ref{sec:sims} each had 100 samples of size 1,000 individuals selected. The run time for a single sample of 1,000 estimating the survey-weighted pseudo-posterior was approximately 17s. Running the na\"\i ve post-hoc adjustment resulted in a total run time of approximately 50s, which includes the initial estimation of the unadjusted pseudo-posterior. This was similar for the prior curvature adjustment. However, the YJ adjustment only added a small additional amount of run time resulting in a total duration of 20s. The YJ adjustment is faster because it avoids estimating the $H_\theta$ matrix on every MCMC draw.

For the NSDUH subsample of size 14,000, the additional time added for the post-hoc YJ adjustment is small. In contrast, running 100 models for the replication approach is two orders of magnitude slower. On the full NSDUH sample, only the YJ adjustment is feasible given time constraints. It still only requires a relatively smaller additional burden compared to the overall run time for estimating the pseudo-posterior model.
\begin{table}
    \centering
    \begin{tabular}{c|c|c|c|c|c}
        Data Set & Size & Warm-up & Draws & Method & run time\\ \hline
        Simulation Sample&  1,000 & 3,000 & 2,000& Unadjusted & 17s\\
         &   & 3,000 & 2,000& Na\"\i ve & 50s\\
         &   & 3,000 & 2,000& Prior Curvature & 50s\\
        &   & 3,000 & 2,000& Yeo-Johnson & 20s\\ \hline
       NSDUH subsample&  14,000 & 3,000 & 2,000& Unadjusted & 7m\\
        &   & 3,000 & 2,000& Yeo-Johnson&  7m 10s\\
        &   & 3,000 & 2,000& Replication & 12h \\ \hline
        Full NSDUH &  236,000 & 3,000 & 2,000& Unadjusted & 3h\\ 
         &   & 3,000 & 2,000& Yeo-Johnson & 3h 10m\\ 
        \hline
    \end{tabular}
    \caption{Approximate run time comparisons for data sets and adjustment approaches for the variance component model \eqref{eq:logisticMM}. Times scale with sample size and the number of Hamiltonian Monte Carlo (HMC) warm-up and sample (draws).}
    \label{tab:runtime}
\end{table}

\end{appendix}

%% Acknowledgements %%
%%%%%%%%%%%%%%%%%%%%%%
%\begin{acknowledgement}%[title={Acknowledgments}]
%\end{acknowledgement}

%% Funding %%
%%%%%%%%%%%%%%%%%%%%%%
%\begin{funding}
%\end{funding}

%\bibliographystyle{jds}  % using the JDS bib style
%\bibliography{biblio}  % biblio.bib should store all your bibtex entries
\bibliography{feb_2025}
\bibliographystyle{chicago}

\end{document}